\documentclass[12pt]{iopart}
\usepackage{iopams}
\usepackage{bm}
\usepackage{amsfonts}
\usepackage{amsthm}
\usepackage{mathtools}
\usepackage{enumerate} 
\usepackage{graphicx}
\usepackage{dcolumn}
\usepackage[utf8]{inputenc}
\usepackage{xcolor}
\usepackage{xspace} 
\usepackage{soul} 
\usepackage[normalem]{ulem} 
\usepackage{hyperref}
\hypersetup{
    colorlinks=true,
    linkcolor=blue,    
    urlcolor=cyan,
    citecolor=blue,
}

\newcommand{\Poincare}{{Poincar\'e}\xspace}
\newcommand{\iotabar}{\mbox{$\iota\!\!$-}}

\newcommand{\bB}{\mathbf{B}}
\newcommand{\bn}{\mathbf{n}}
\newcommand{\matT}{\mathcal{T}}

\newcommand{\real}{\mathbb{R}}

\newcommand{\naturalnum}{\mathbb{N}}

\newcommand{\trtrace }{\mbox{Tr}}
\newcommand{\lcontfrac}{\langle}
\newcommand{\rcontfrac}{\rangle}

\renewcommand{\pt}{p_\theta}
\newcommand{\pz}{p_\zeta}
\newcommand{\gtt}{g^{\theta \theta}}
\newcommand{\gtz}{g^{\theta \zeta}}
\newcommand{\gzz}{g^{\zeta \zeta}}

\newcommand{\figref}[1]{{Figure \ref{#1}}}
\newcommand{\Figref}[1]{{Figure \ref{#1}}}
\newcommand{\secref}[1]{{Section \ref{#1}}}
\newcommand{\Eqref}[1]{{Equation \eqref{#1}}}

\newcommand{\modi}{\color{black}}
\newcommand{\modii}{\color{black}}
\newcommand{\norm}{\color{black}}

\definecolor{red}{rgb}{1,0,0}
\definecolor{gre}{rgb}{0,1,0}
\definecolor{blu}{rgb}{0,0,1}

\newcommand{\ZSQ}[1]{}

\renewcommand{\etal}{{\em et al. }}
 
\begin{document}

\bibliographystyle{Science}

\title{On the non-existence of stepped-pressure equilibria far from symmetry}

\author{Z. S. Qu$^1$, S. R. Hudson$^2$, R. L. Dewar$^1$, J. Loizu$^3$, M. J. Hole$^{1,4}$}

\address{$^1$Mathematical Sciences Institute, the Australian National University, Canberra ACT 2601, Australia}
\address{$^2$Princeton Plasma Physics Laboratory, PO Box 451, Princeton, New Jersey 08543, USA}
\address{
$^3$\'Ecole Polytechnique F\'ed\'erale de Lausanne, Swiss Plasma Center, CH-1015 Lausanne, Switzerland
}
\address{$^4$Australian Nuclear Science and Technology Organisation, Locked Bag 2001, Kirrawee DC NSW 2232, Australia}
\ead{zhisong.qu@anu.edu.au}
\vspace{10pt}

\begin{abstract}
The Stepped Pressure Equilibrium Code (SPEC) [Hudson \etal, Phys. Plasmas 19, 112502 (2012)] has been successful in the construction of equilibria in 3D configurations
that contain a mixture of flux surfaces, islands and \modi chaotic magnetic field lines. \norm
\modi In this model, \norm
the plasma is sliced into sub-volumes separated by ideal interfaces, \modi and in each volume the magnetic field is a Beltrami field. \norm 
In the cases where the system is far from possessing a continuous symmetry, such as in stellarators, 
the existence of solutions to a stepped-pressure equilibrium with given constraints, such as a multi-region relaxed MHD minimum energy state, 
is not guaranteed but is often taken for granted.
Using SPEC, we have studied two different scenarios in which a solution 
fails to exist in a slab with \modi analytic \norm boundary perturbations.
We found that with a large boundary perturbation, a certain interface becomes \modi fractal, \norm
corresponding to the break up of a Kolmogorov–Arnold–Moser (KAM) surface.
Moreover, an interface can only support a maximum pressure jump 
\modi while a solution of the magnetic field consistent with the force balance condition can be found. \norm
An interface closer to break-up can support \modii a smaller pressure jump. \norm
We discovered that the pressure jump can push the interface closer to being non-smooth through force balance,
thus significantly decreasing the maximum pressure it can support.
Our work shows that a convergence study must be performed on a SPEC equilibrium with interfaces close to break-up.
These results may also provide insights into the choice of interfaces and have applications in finding out the maximum pressure a machine can support.
\end{abstract}

%
%
\submitto{\PPCF}
%
%
%

\section{Introduction}

An isotropic, static ideal magneto-hydrodynamics (MHD) equilibrium is described by the force balance equation
\begin{equation}
  \nabla p = \mathbf{J} \times \bB,
  \label{eq:force_balance_MHD}
\end{equation}
where $p$ is the plasma pressure, $\bB$ the magnetic field, 
$\mathbf{J} = \nabla \times \bB / \mu_0$ the current density, and $\mu_0$ the magnetic permeability constant.
Finding such an equilibrium in \modi three-dimensional \norm (3D) stellarator geometry is a complicated problem~\cite{Grad1967}.
\modi
Taking the dot product of \eqref{eq:force_balance_MHD} and $\bB$ gives the magnetic differential equation $\bB \cdot \nabla p = 0$,
meaning that the pressure is a constant along a field line.
\norm
Solving \eqref{eq:force_balance_MHD} thus requires resolving the structure of magnetic field lines
which are unknown \textit{a priori}.

The magnetic field lines are constructed by integrating $d\bm{x}/dt =\bB(\bm{x})$,
in which $\bm{x}$ is the coordinate of a point on the field line and $t$ a time-like variable specifying the distance along a field line. 
\modi
This gives a $1 \frac{1}{2}$ degrees of freedom Hamiltonian system \cite{Meiss1990} since the magnetic field is divergence-free, and the physical quantities have the periodicity naturally associated with the toroidal geometry.
\norm
With a loss of symmetry, the Hamiltonian systems with more than one dimension are in general not integrable, i.e. flux surfaces are not guaranteed to exist. 
When the departure from symmetry is small most flux surfaces with a sufficiently irrational rotational transform $\iotabar$ still persist, 
as suggested\footnote{Except for vacuum fields with fixed boundaries, standard finite-dimensional KAM theory is simply inapplicable to magnetic field-line flow in a plasma, even though the flow is Hamiltonian.
This is because its $1 \frac{1}{2}$-d.o.f. Hamiltonian cannot be put in the standard form $H_0 + \epsilon H_1$, with $H_0$ and $H_1$ known and independent of $\epsilon$, because this field-line Hamiltonian is a  functional of the plasma currents, which change in a complicated way, to be determined self-consistently as part of  the problem. It is true ideal MHD is Hamiltonian, but it is infinite-dimensional so standard KAM theory cannot be applied to this problem.}
by the Kolmogorov-Arnold-Moser (KAM) theorem \cite{Arnold1963}.
These flux surfaces are also called KAM surfaces, 
\modi or invariant tori because the field-line flow maps  such a surface onto itself. \norm
However, magnetic islands will open up on every rational flux surface. 
Chaos will develop around islands with the field lines exhibiting 
\modi a random-looking, locally ergodic behaviour. 
When the departure from symmetry becomes larger, islands will begin to overlap and smooth KAM surfaces will become increasingly ``wrinkly'' until, just before  breaking up, they become fractal invariant tori having a hierarchy of qualitatively self-similar structure on all scales. After breakup, the invariant set is no longer connected into a torus, but rather becomes a disconnected  ``cantorus'', still a transport barrier but leaky. That is, if followed  sufficiently far a field line may eventually find a gap in the cantorus that allows it to cross to the other side of the original invariant torus. \norm 

As flux surfaces, islands and chaos coexist in 3D magnetic fields,
a \modi nontrivial \norm pressure profile consistent with these structures is  complicated.
\modi
In ideal MHD, which ``freezes in'' the topology of the magnetic field,  one can constrain the class of solutions to have nested toroidal flux surfaces everywhere and use an energy minimization method based on the MHD variational principle first put forward by Kruskal and Kulsrud~\cite{Kruskal1958}. This approach was adopted in the well-known Variational Moments Equilibrium Code (VMEC)~\cite{Hirshman1986}.
\norm
However, on flux surfaces with a rational rotational transform and a non-zero pressure gradient,
there is an unphysical $1/x$ Pfirsch-Schl\"{u}ter current density whose surface integral is not bounded~\cite{Loizu2015b,Loizu2015a}.
The \modi generic \norm presence of islands and chaos is also ignored.
To avoid these problems, many codes such as PIES~\cite{Reiman1986}, HINT~\cite{Suzuki2006} and SIESTA~\cite{Hirshman2011}
take into account non-ideal effects and perturbations, but are usually more computationally expensive. 
The third approach is to relax the smoothness condition of the magnetic field and allow discontinuities.
This leads to the stepped-pressure equilibrium and will be the focus of the current paper.

In a stepped-pressure equilibrium~\cite{Hole2007, Hole2009, Hudson2012b},
the plasma volume is partitioned into a number of sub-volumes with non-relaxing ideal interfaces between them.
Within each volume, the magnetic field satisfies a Beltrami equation.
The pressure $p$ is a constant within each volume but can jump at the interfaces,
leading to ``stepped-pressure equilibrium'' in its name.
\modi The magnetic energy density $B^2/(2\mu_0)$, which also acts as a pressure exerting a force on interfaces, has a counterbalancing jump at these interfaces so the total pressure, kinetic plus magnetic, is continuous.\norm
A stepped-pressure equilibrium can be formulated as a stationary point of the Multi-region Rela\underline{x}ed MHD (MRxMHD) energy functional ~\cite{Hole2007, Hole2009}.
Physically, this is equivalent to the plasma undergoing a Taylor relaxation \cite{Taylor1974,Taylor1986} within each sub-volume,
in which the total energy is minimized subject to the constraint of conserved magnetic helicity and magnetic fluxes.

To access stepped pressure equilibrium solutions numerically, 
the Stepped-Pressure Equilibrium Code (SPEC) ~\cite{Hudson2012b, Hudson2012a} was built \modi in 2012 \norm.
Since then, the MRxMHD model and SPEC have led to a number of interesting and important findings,
such as the equilibrium bifurcation in reversed-field pinch plasmas~\cite{Dennis2013a},
the formation of singular current sheets in 3D equilibria~\cite{Loizu2015b},
the penetration of resonant magnetic perturbations ~\cite{Loizu2015a,Loizu2016b},
the equilibrium beta limit~\cite{Loizu2017} and saturation of tearing modes~\cite{Loizu2019, Loizu2020}.
MRxMHD and SPEC have been extended in many different ways to calculate free-boundary equilibria~\cite{Hudson2020} and equilibria with fixed current~\cite{Baillod2021},
and to include the effect of flow~\cite{Dennis2014a,Qu2020},
pressure anisotropy ~\cite{Dennis2014b}, two fluid effects~\cite{Lingam2016} and
time evolution~\cite{Dewar2015,Dewar2017,Dewar2020}.
SPEC is also optimised in terms of its speed and robustness \cite{Qu2020b}.
Apart from SPEC, a new code, the Boundary Integral Equation Solver for Taylor states (BIEST)~\cite{ONeil2018, Malhotra2019}, 
was developed recently based on the same stepped-pressure equilibrium 
but using a boundary integral method to solve the Beltrami equation 
\modi rather than the spectral Galerkin method used in SPEC within each volume. \norm

A problem often overlooked is how to select and where to place the ideal interfaces.
Physically, one may choose to place a large number of interfaces at locations where there is a large pressure gradient as measured experimentally,
and place fewer interfaces at the locations where the pressure is flat.
Mathematically, a solution to such choices of interfaces and discrete profiles (pressure, fluxes, interface $\iotabar$, helicity, Beltrami parameter $\mu_i$, etc.) should exist \modii for the numerical calculation to be meaningful. \norm
Bruno and Lawrence~\cite{Bruno1996} proved the existence of stepped-pressure equilibria when the departure from axisymmetry is small and with the interfaces requiring a sufficiently irrational $\iotabar$.
However flux surfaces in stellarators, the main target application of SPEC, cannot be considered as small perturbations away from axisymmetric tori:
\modi
their strong shaping has a dominant effect in determining the magnetic field inside the plasma,
in particular the rotational transform.
\norm
There is no guarantee to have a solution when the departure from symmetry is large, 
although the existence is often taken for granted 
\modi
without further verification.
\norm
Moreover, cases in which a solution does not exist are seldom demonstrated or studied.

In this paper, we aim to show numerically using SPEC the transition from existence to non-existence of solutions.
It will help to identify regimes for which a pressure-supporting flux surface is broken,
which then leads to an enhanced transport in its vicinity.
This will also provide insights into the selection of interfaces as a trial and error process:
a lack of solution often indicates that one should consider removing the interface and merging the two neighbouring volumes.
\secref{sec:preliminaries} formally introduces the stepped-pressure equilibrium and
summarises the prerequisites: the irrationality and the analyticity of the interfaces.
The section also introduces the Pressure Jump Hamiltonian (PJH) \cite{McGann2010,McGann2013} as a tool to find the magnetic field on an interface that is consistent with the force balance condition.
In \secref{sec:kam_surface_break_up}, we set up our problem using a 
\modi
Hahm-Kulsrud-Taylor slab \cite{Hahm1985}
\norm
with boundary perturbations of multiple Fourier harmonics.
\modi
We attempt to place an interface on the remnant cantorus after a KAM surface is broken,
leading to an interface with discontinuous first derivatives, signalling the non-existence of such an equilibrium solution.
\norm
In the second scenario, a finite pressure jump on the interface could lead to the non-existence of equilibrium solutions.
We found that a single SPEC run gives no indication of the non-existence straight away
\modi unless the Fourier resolution is extremely high\norm.
Rather, the non-existence can manifest itself as a lack of convergence as the numerical resolution increases.
This is the theme of \secref{sec:existence_PJH}, in which a convergence criterion based on the PJH return map is developed to find the range of pressure jump values an interface can endure.
Finally, \secref{sec:conclusion} discusses the results and draws conclusions.

\section{Preliminaries}
\label{sec:preliminaries}
\subsection{The stepped-pressure equilibrium}
In the stepped-pressure equilibrium,
the plasma volume is partitioned into a number of sub-volumes.
Within each volume, the magnetic field satisfies the Beltrami equation
\begin{equation}
    \nabla \times \bB = \mu_i \bB,
    \label{eq:Beltrami}
\end{equation}
where $\mu_i$ is referred to as the helicity multiplier in the $i$-th volume.
These sub-volumes are separated by \textit{ideal interfaces} with the boundary condition given by
\begin{equation}
    \bB \cdot \mathbf{n} = 0,
    \label{eq:ideal_boundary_condition}
\end{equation}
where $\mathbf{n}$ is the unit vector normal to the surface of the interface.
\modi
This implies that the interface should have continuous first derivatives, i.e. $C^1$, 
since the normal vector $\bn$ will be ill-defined otherwise.
\modii
\Eqref{eq:ideal_boundary_condition} can be extended in a weaker sense to incorporate interfaces with a finite number of break points, e.g. an interface with an X point.
It is satisfied approaching the corner from both sides, but not exactly on the corner.
\norm
\norm

The total pressure (thermal plus magnetic) is balanced on the interfaces, giving that
\begin{equation}
    \left[\left[ p + \frac{B^2}{2} \right]\right]=0,
    \label{eq:force_balance}
\end{equation}
where the jump operator $[[\cdots]]$ stands for the difference between either side of the interface.
\modi The magnetic field strength $B$ here should be understood as $B / \sqrt{\mu_0}$ in SI units. \norm

Apart from pressure, one needs to specify three constraints for each sub-volume (except the innermost volume of a cylinder or toroid, which needs only two).
In the original MRxMHD theory, the toroidal flux $\psi_{t,i}$, the poloidal flux $\psi_{p,i}$ and the magnetic helicity 
are specified.
However, the helicity is sometimes an inconvenient quantity 
\modi as \norm one loses control of the $\iotabar$ profile and/or the current profile.
Instead, for an equilibrium code one can specify $\iotabar$ on the inner/outer side of each interface, or specify the current on each interface and within each volume~\cite{Baillod2021}.
The poloidal flux $\psi_{p,i}$ and helicity multiplier $\mu_i$ are iterated to satisfy these constraints.
The helicity in each volume can be computed \textit{a posteriori}.
In this paper, we will prescribe a discrete toroidal flux and $\iotabar$ profile.

In the numerical scheme of SPEC, one first provides a guess of \modi the location of the interfaces. \norm
The Beltrami equation \eqref{eq:Beltrami} is then solved with the boundary condition \eqref{eq:ideal_boundary_condition} and aforementioned volume-wise constraints satisfied.
SPEC uses a spectral Galerkin method \cite{Qu2020b}~:
the field in each volume is discretised into a combination of Chebyshev polynomials in its radial direction and Fourier series in its two angles.
The highest mode number of Chebyshev or Fourier basis functions in the radial, poloidal and toroidal directions are denoted by $L$, $M$ and $N$, respectively. 
Using the obtained magnetic field, the force difference on each interface, i.e. the left hand side of \eqref{eq:force_balance}, is computed.
A nonlinear root-finding algorithm then iterates on the interface geometry (specified by the Fourier coefficients) to reduce the force difference.
Within each iteration step, the magnetic field is recomputed using the updated guess of the interfaces.
The process is repeated until force balance \eqref{eq:force_balance} is satisfied.
The output of SPEC contains two parts: the magnetic field in each volume, and the geometry of the interfaces separating them.

\subsection{The pressure jump Hamiltonian}
\label{sec:preliminaries_PJH}
If the interface geometry, the total magnetic field strength $B$ on one side of the interface and the thermal pressure difference $\Delta p$ are known,
the force balance condition can be used to derive the magnetic field on the other side of the interface.
Berk \etal~\cite{Berk1986} were the first to formulate it as a Hamilton-Jacobi equation,
and then investigated the possibility of using
\modi
KAM theory 
\norm
to infer the existence of equilibria with sharp boundaries.
Kaiser and Salat~\cite{Kaiser1994} followed a different route by converting the problem into finding a covering of geodesics on the surface.
McGann \etal~\cite{McGann2013} extended the work of Berk \etal for plasma-vacuum system to the class of stepped-pressure equilibria.
In this paper, we will follow the 
\modi
formalism
\norm
developed by McGann \etal.

The condition \eqref{eq:force_balance} can be rewritten as
\begin{equation}
  2 (p^- - p^+)  = 2 \Delta p= (B^{+})^2 - (B^{-})^2,
  \label{eq:pressure_jump}
\end{equation}
in which the superscripts $+$ and $-$ label the outer and inner side of the interface, respectively.
Let the interface be described by two periodic angles $\theta$ and $\zeta$ labeling poloidal and toroidal directions.
This interface is a two-dimensional Riemannian manifold with covariant metric $g_{ij}$, where $i,j \in \{\theta,\zeta\}$.
The contravariant metric $g^{ij}$ can be obtained by inverting the $2 \times 2$ covariant metric tensor.
Given \eqref{eq:ideal_boundary_condition}, one gets
\begin{equation}
  (B^{\pm })^2 = \sum_{i,j \in \{\theta,\zeta\}} g^{ij} B^{\pm }_i B^{\pm }_j, 
  \label{eq:B2_covariant}
\end{equation}
in which $B^{\pm }_i$ is the covariant component of $\bB^{\pm }$ on the outer/inner side of the interface.

According to the Beltrami equation \eqref{eq:Beltrami}, the current is parallel to the magnetic field and therefore on the interfaces,
$(\nabla \times \bB^{\pm })  \cdot \bn = 0$.
This can be rewritten as
\begin{equation}
  \partial_{\theta} B^{\pm }_\zeta - \partial_\zeta B^{\pm }_\theta = 0.
  \label{eq:current_boundary_condition}
\end{equation}
\Eqref{eq:current_boundary_condition} is 
\modi
automatically
\norm
satisfied if the field components are written as
\begin{align}
  B^{\pm}_\theta = \partial_\theta f^{\pm}, \quad B^{\pm}_\zeta = \partial_\zeta f^{\pm},
  \label{eq:surface_potential}
\end{align}
where the two scalar functions $f^{\pm}(\theta, \zeta)$ are known
\modi
as
\norm
\textit{surface potentials}
which define the field components on the outer/inner side of the interface.

Without loss of generality, we can take $(B^-)^2$ as known and $(B^+)^2$ as unknown.
Substituting \eqref{eq:surface_potential} into \eqref{eq:B2_covariant} and then \eqref{eq:pressure_jump} will give
\modi
\begin{align}
  H(\theta, \zeta, \partial_\theta f^+, \partial_\zeta f^+) = 2 \Delta p = \sum_{i,j \in \{\theta,\zeta\}} g^{ij} \partial_i f^+ \partial_j f^+ - V(\theta, \zeta),
\end{align}
\norm
in which $V(\theta, \zeta) = (B^{-})^2$.
This is the Hamilton-Jacobi equation for the time-independent Hamiltonian
\begin{align}
  H(\theta, \zeta, \pt, \pz) = 2 \Delta p = \sum_{i,j \in \{\theta,\zeta\}} g^{ij} p_i p_j - V(\theta, \zeta),
  \label{eq:PJH}
\end{align}
with $\pt = \partial_\theta f^+$ and $\pz = \partial_\zeta f^+$ and $f^+$ being the generating function.
\Eqref{eq:PJH} is known as the \textit{pressure jump Hamiltonian}(PJH) .
The equations of motion in the phase space of the PJH are given by the standard Hamilton's equations.
This is a system of ordinary differential equations for the momenta $\pt$ and $\pz$,
and the coordinates $\theta$ and $\zeta$, 
as a function of  a time-like independent variable $t$.

\Eqref{eq:PJH} can be reduced to a $1 \frac{1}{2}$ degree-of-freedom system by noting that the angle $\zeta$ is monotonically increasing with $t$ \modi if the toroidal field does not change sign on the interface \norm,
and can be used to replace $t$.
The equations of motion for $\theta$ and $\pt$ as a function of $\zeta$ are given by
\begin{align}
  \label{eq:PJH_dtheta}
  \frac{d \theta}{d \zeta} &= \frac{\partial_{\pt}H}{\partial_{\pz}H} = \frac{\gtt \pt + \gtz \pz}{\gtz \pt + \gzz \pz} = \frac{B^{\theta+}}{B^{\zeta+}}, \\
  \label{eq:PJH_dptheta}
  \frac{d p_\theta}{d \zeta} &= - \frac{\partial_\theta H}{\partial_{\pz}H} = 
  - \frac{ \partial_\theta (\gtt \pt^2 + 2 \gtz \pt \pz + \gzz \pz^2 - V)}{2 (\gtz \pt + \gzz \pz) },
\end{align}
in which $\pz = \pz(\theta, \zeta, \pt; \Delta p)$ is computed by inverting \eqref{eq:PJH} given $\Delta p$.
This is because the Hamiltonian \eqref{eq:PJH} does not depend explicitly on time,
and the total ``energy'' is a conserved quantity.
It is noteworthy that \eqref{eq:PJH_dtheta} is the same as the equation for magnetic field lines \eqref{eq:field_line_flow}
in \ref{app:greenes_residue}.
The definition of rotational transform on the ``$+$'' side, given by
\begin{equation}
  \iotabar^+ \equiv \lim_{\zeta \rightarrow \infty} \frac{ \theta(\zeta) - \theta_0}{\zeta - \zeta_0},
\end{equation}
is thus shared between the PJH and the magnetic field lines,
in which $\theta(\zeta) $ is obtained by solving \eqref{eq:PJH_dtheta} and \eqref{eq:PJH_dptheta}
with initial values $\theta(\zeta_0) = \theta_0$ and $\pt(\theta_0) = p_{\theta,0}$.
In other words, if a solution of the PJH has some $\iotabar^+$ in the PJH phase space,
the corresponding field line in real space has the same $\iotabar^+$.
\modi A solution to the Hamilton-Jacobi equation corresponds to a KAM surface in the phase space of the Hamiltonian. \norm
Consequently, finding the magnetic field with a given $\iotabar^+$ satisfying the force balance equation \eqref{eq:force_balance} is converted into finding a KAM surface with the same $\iotabar^+$ in the phase space of PJH.
McGann \etal \cite{McGann2010}~ found a one to one relationship between a KAM surface in the PJH phase space and a solution of $\bB^+$ satisfying the force balance condition.
In other words, if one can find such a KAM surface in the PJH phase space, there is also a corresponding solution of $\bB^+$ satisfying the force balance in real space.
On the contrary, if such a KAM surface does not exist, there is no corresponding solution of $\bB^+$ satisfying the force balance condition.

\subsection{Irrationality condition}
\label{sec:analyticity_irrationality}

In this paper, we will only consider the class of equilibrium with the same irrational $\iotabar$ on both the inner and outer side of the interface.
We will stick to the original picture of MRxMHD proposed by Hudson \etal \cite{Hudson2012b}, 
that the interfaces are pressure-supporting surfaces playing the same role as the KAM surfaces.
Since all the KAM surfaces have irrational rotational transforms, so do the interfaces.

\section{Non-existence due to KAM surfaces breaking up}
\label{sec:kam_surface_break_up}
\subsection{A single-volume Hahm-Kulsrud-Talyor slab with emerging chaos}
\label{sec:one_volume}
In this section, we start our investigation in slab geometry and introduce a modified version of the Hahm-Kulsrud-Talyor (HKT) problem~\cite{Hahm1985}.
It is the simplest problem that maintains all the features of flux surfaces, islands and chaos. 
Consider a MRxMHD equilibrium with one volume and zero pressure in a slab.
To mimic the toroidal geometry, we define the $y$ and $z$ coordinates to be periodic in $2\pi$
and introduce the ``poloidal'' and ``toroidal'' angles $\theta = y$ and $\zeta = z$.
The ``radial'' coordinate is defined to be $R = x$, in which $(x,y,z)$ are the Cartesian coordinates.
In the non-perturbed case, two ideal walls are placed at $R_{\text{down}}=0$ and $R_{\text{up}}=1$ with the plasma sitting between the walls.

\begin{figure*}[!htbp]
  \[
  \begin{array}{c c}
    \includegraphics[width=8cm]{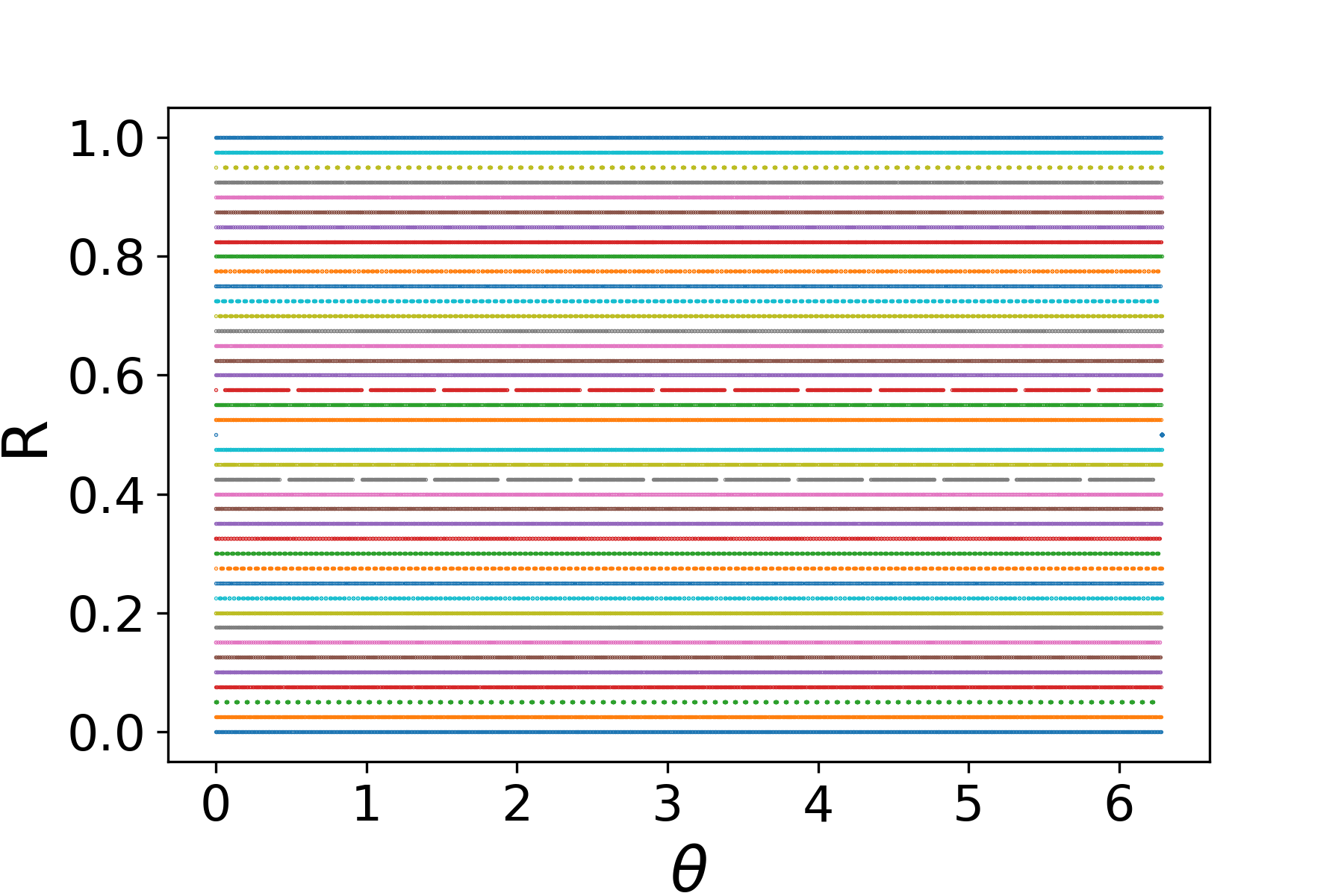} &
    \includegraphics[width=8cm]{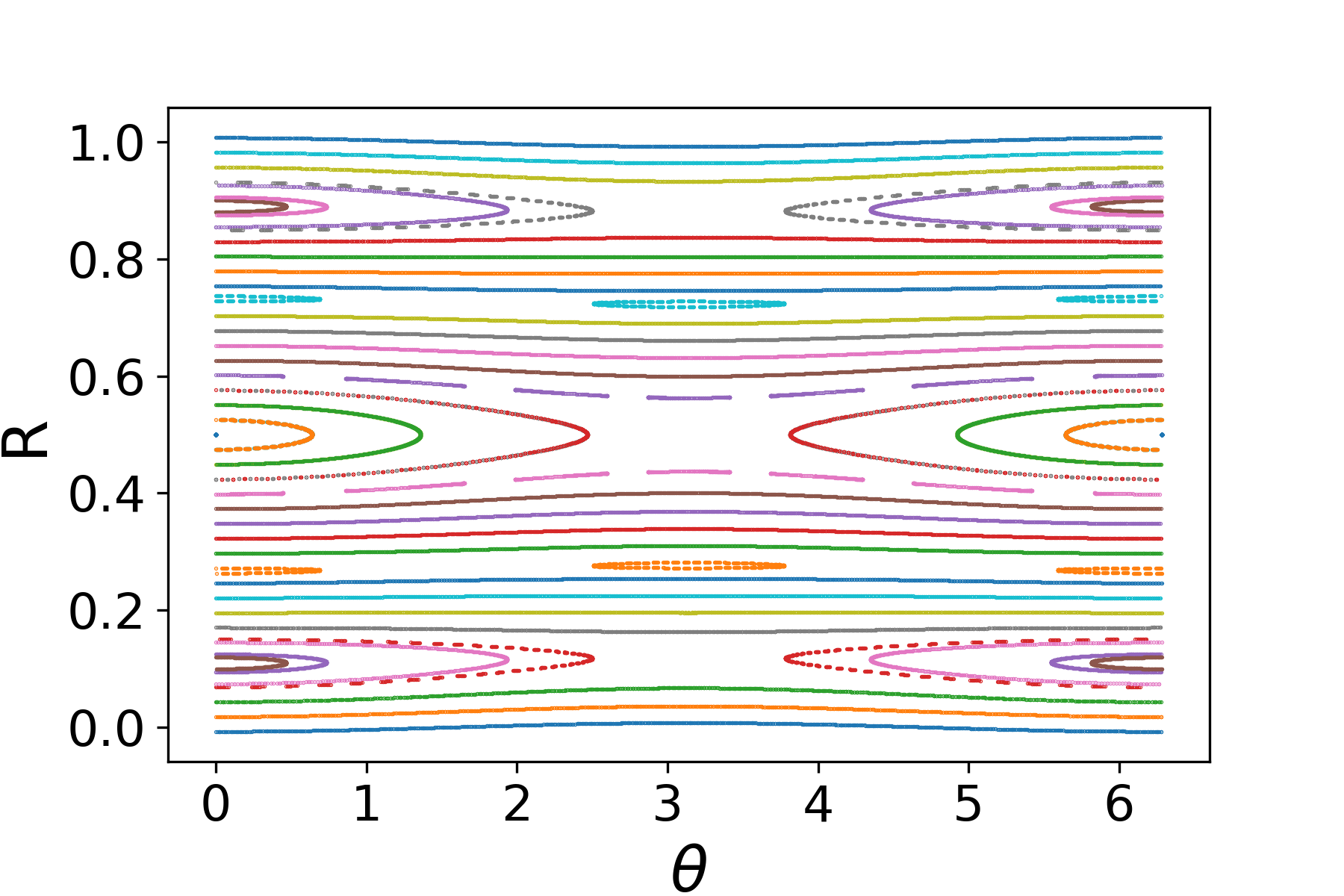} \\
    \text{(a)} & \text{(b)} \\
    \includegraphics[width=8cm]{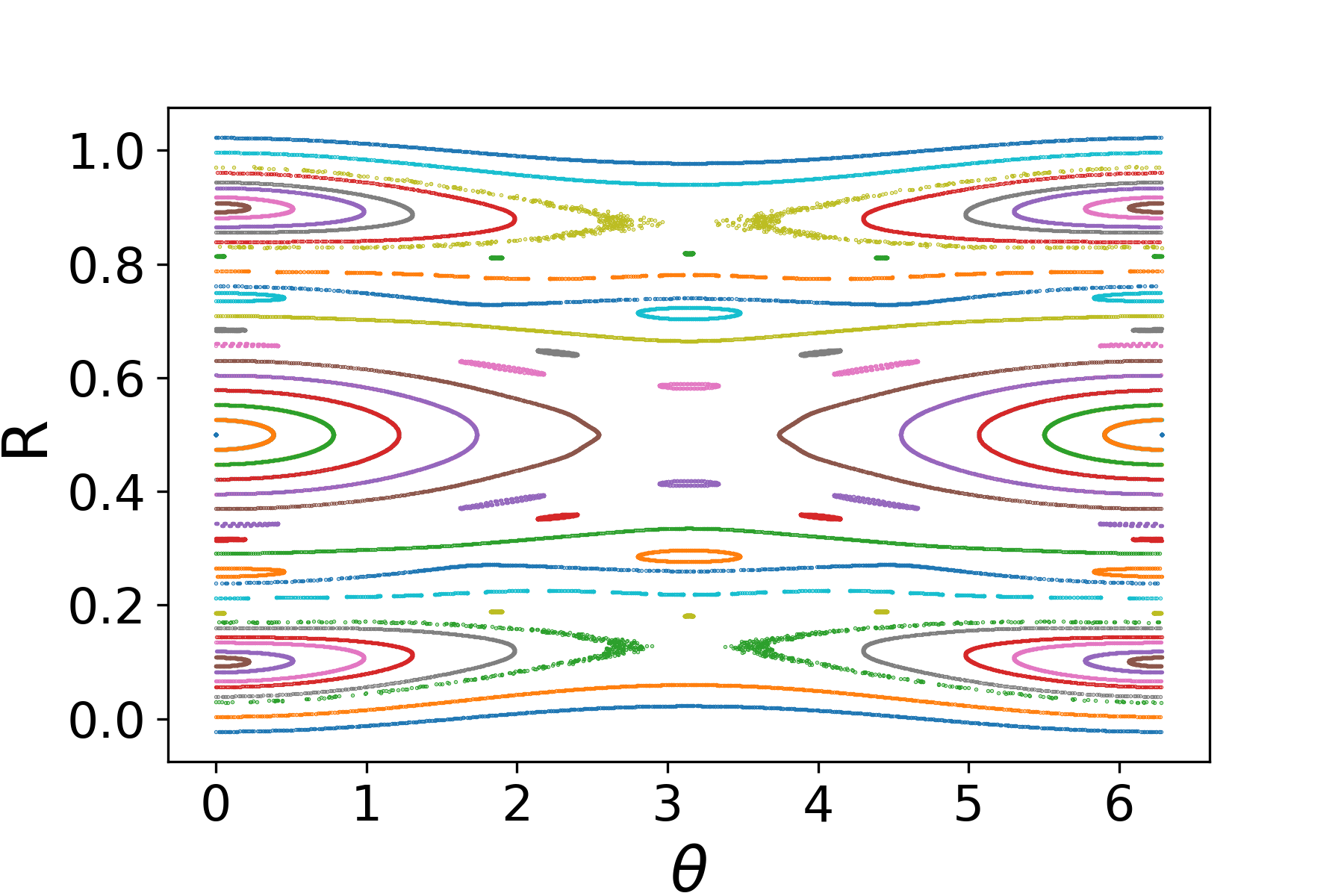} &
    \includegraphics[width=8cm]{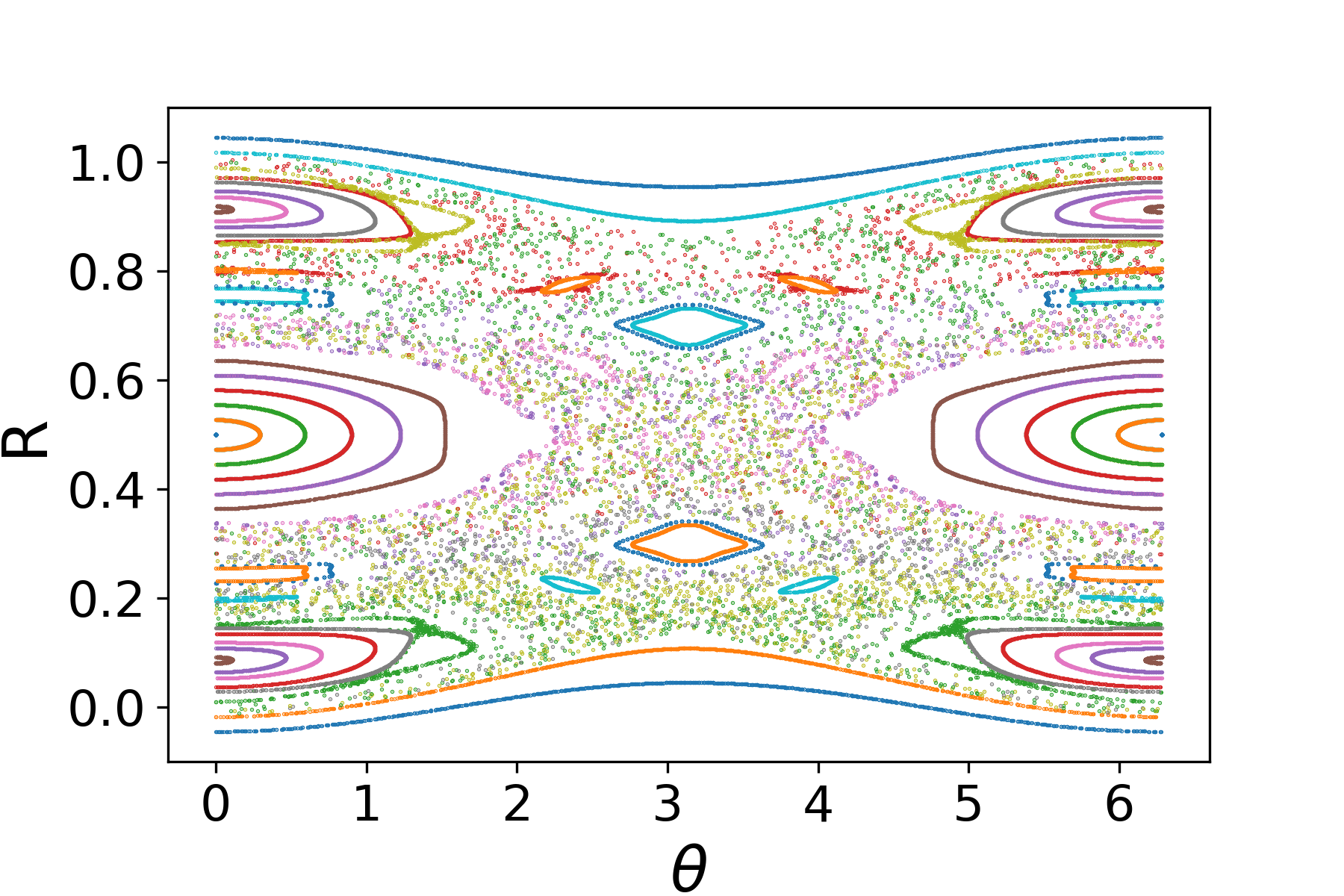} \\
    \text{(c)} & \text{(d)}
  \end{array}
  \]
  \caption{The \Poincare plot of the single-volume HKT problem at $\zeta = 0$ with (a) $\delta=0$ (b) $\delta = 0.005$ (c) $\delta = 0.015$ and (d) $\delta=0.03$.}
  \label{fig:poincare_1vol}
\end{figure*}

\begin{figure}[!htbp]
  \centering
  \includegraphics[width=8cm]{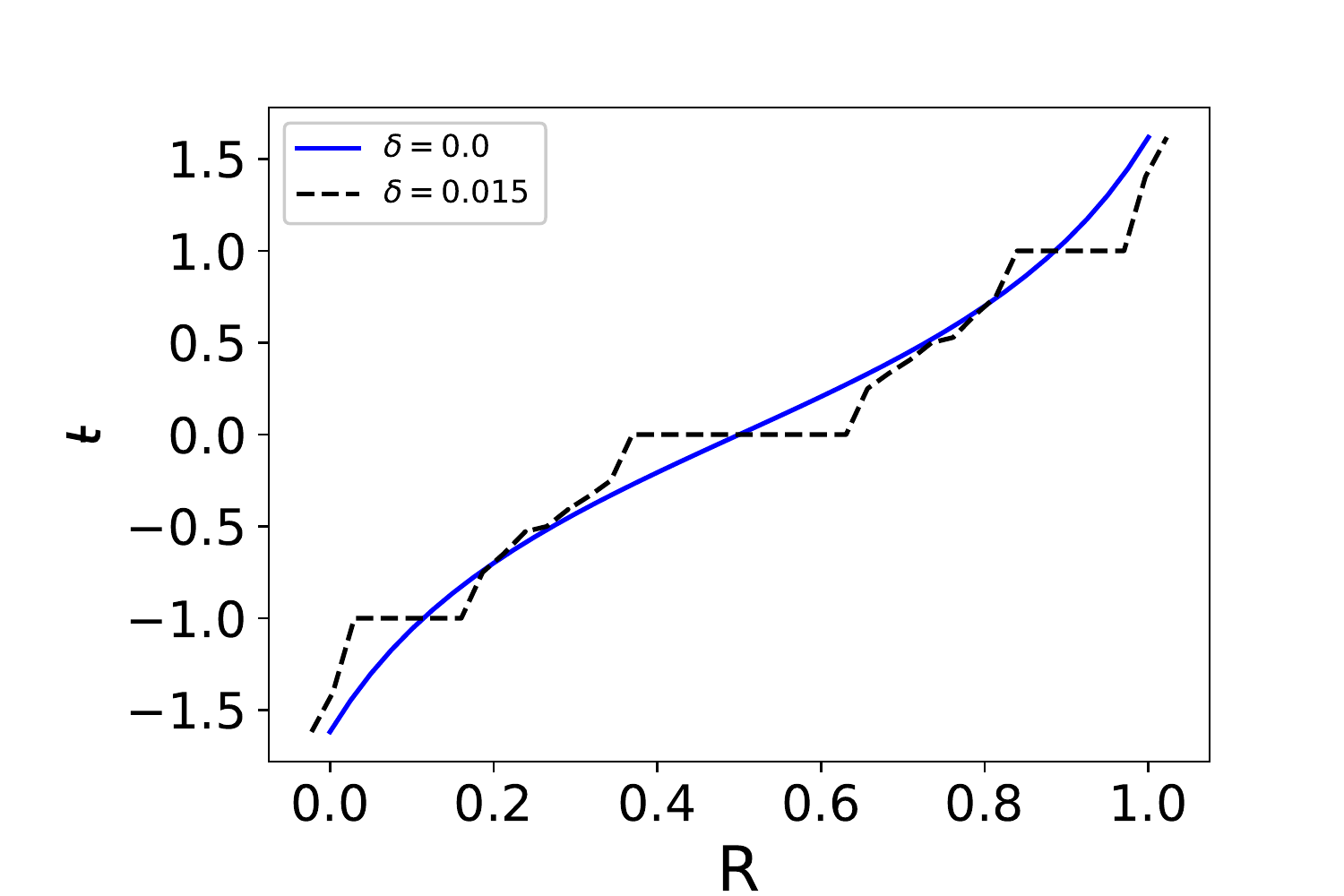}
  \caption{The rotational transform $\iotabar$ profile, evaluated by tracing field lines starting at different values of $R$ and $\theta=\zeta=0$.}
  \label{fig:iota_1vol}
\end{figure}

The magnetic field in the plasma volume follows the Beltrami equation $\nabla \times \bB = \mu \bB$.
\modi
Considering the symmetry over the $R=1/2$ plane, the solution $\bB_0$ is given by
\norm
\begin{equation}
    B_{0,R}(R) = 0, \quad B_{0,\theta}(R) = B_0 \sin \mu \left(R - \frac{1}{2}\right), \quad B_{0,\zeta}(R) = B_0 \cos \mu \left(R - \frac{1}{2}\right),
\end{equation}
in which $B_0 = |\bB_0| = \mu / (4 \pi \sin (\mu/2) )$ is a constant that normalizes the toroidal flux to unity.
The toroidal flux is defined by
\begin{equation}
  \psi_t = \int_{0}^{1} \int_{0}^{2 \pi} \bB(R,\theta, \zeta=0) \cdot \mathbf{e}_\zeta dR d\theta,
\end{equation}
on the $\zeta = 0$ plane, in which $\mathbf{e}_\zeta$ is the unit vector in $\zeta$ direction.
The rotational transform in the non-perturbed case is given by
\begin{equation}
    \iotabar(R) = \tan \left[\mu \left(R - \frac{1}{2}\right)\right],
\end{equation}
with the $\iotabar = 0$ surface sitting at $R = 1/2$.
This solution satisfies the boundary condition $\bB \cdot \bn =0$.
\Figref{fig:poincare_1vol} gives the \Poincare section at $\zeta = 0$ and \figref{fig:iota_1vol} gives the profile of \iotabar.
\modi
In all the cases below, we have chosen the smallest positive $\mu$ such that $\iotabar$ on the lower and upper wall are given by $\iotabar^-=-\varphi$ and $\iotabar^+=\varphi$, respectively,
\norm
where
\begin{equation}
    \varphi = \frac{1+\sqrt{5}}{2} \approx 1.618 \cdots,
\end{equation}
is the golden ratio.

Next, we consider the case where the upper boundary is perturbed to
\begin{align}
    R_{\text{up}}(\theta, \zeta)= 1 + \frac{\delta}{2} [\cos(\theta - \zeta) +  \cos(\theta) + \cos(\theta + \zeta)],
\end{align}
and the lower boundary symmetrically to
\begin{align}
    R_{\text{down}}(\theta, \zeta)=  - \frac{\delta}{2} [\cos(\theta - \zeta) +  \cos(\theta) + \cos(\theta + \zeta)].
\end{align}
Unlike the original HKT problem in which the boundary is perturbed by a single Fourier harmonic,
we perturb the boundary with $m=1,n=0,\pm 1$ modes, in which $m$ and $n$ are the poloidal and toroidal mode numbers.
Consequently, islands develop at $\iotabar = 0, \pm 1$ and all other rational surfaces between them due to the nonlinear interaction of the main islands,
with many primary flux surfaces in the rest of the plasma volume.
This is shown in \figref{fig:poincare_1vol} (b) and (c).
The existence of such surviving flux surfaces is suggested, \modi for sufficiently small $\delta$, \norm by 
the KAM theorem~\cite{Arnold1963},
which says that the invariant tori (also called flux surfaces or KAM surfaces) with a Diophantine 
(a class of irrationals that are far enough from rationals)
rotational transform will survive a small enough symmetry-breaking perturbation.
If we keep increasing $\delta$, 
the islands continue to grow,
destroying the KAM surfaces between them.
Chaos will develop from the X points of the islands.
The corresponding $\iotabar$ profile in \figref{fig:iota_1vol} 
computed by field line tracing, has flattened regions at the location of the islands.
When $\delta$ is large enough, the last KAM surface will be broken, as shown by \figref{fig:poincare_1vol} (d).
The pictures here are analogous to those of the standard map~\cite{Meiss1992}.

\begin{figure}[!htbp]
  \centering
  \includegraphics[width=8cm]{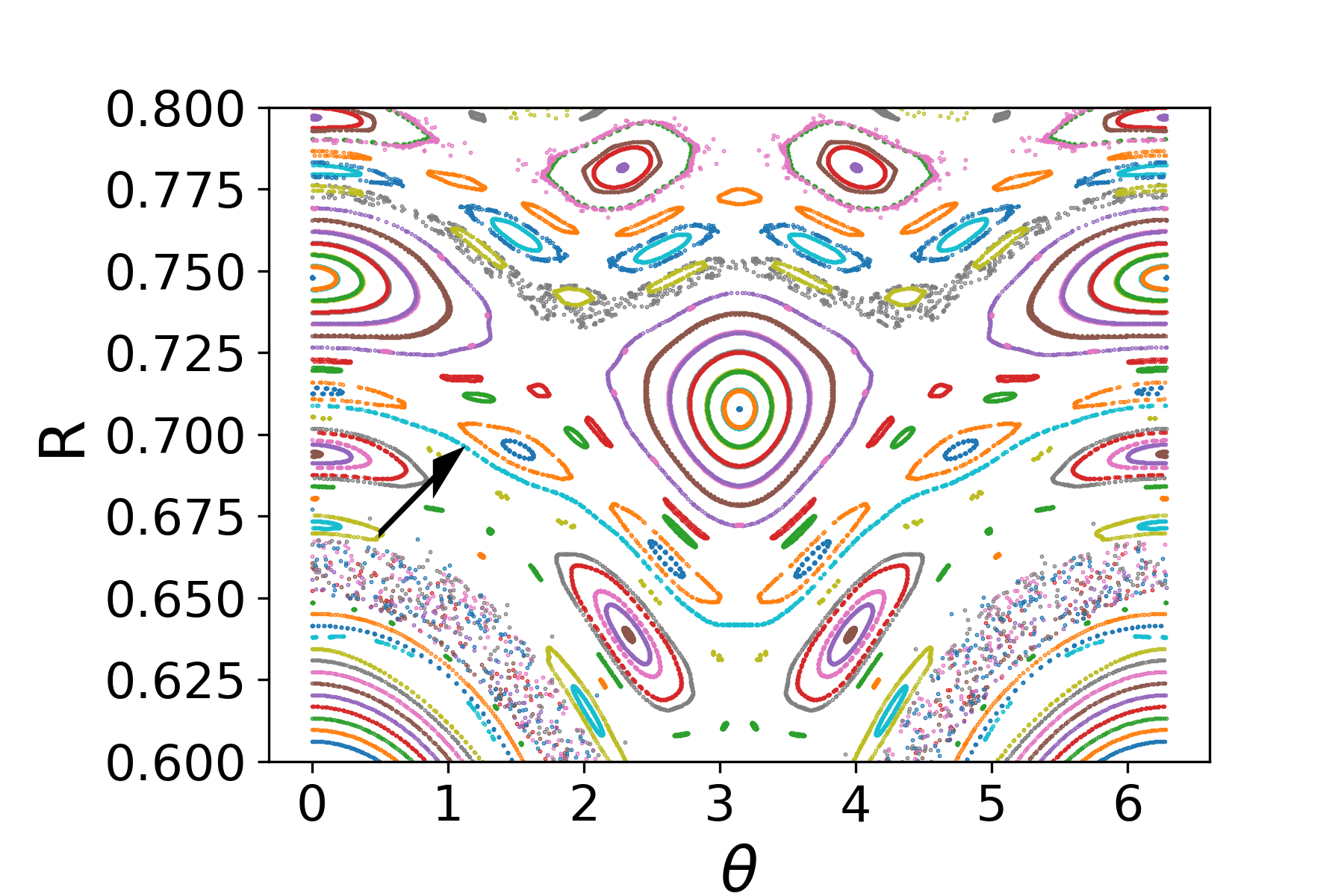}
  \caption{The \Poincare plot of the single-volume HKT slab with $\delta=\delta_c =0.020782$, 
  zoomed into the region $0.6 < R < 0.8$ around the last standing KAM surface passing through $R=0.71$ and $\theta=0$ as indicated by the arrow. }
  \label{fig:poincare_zoom}
\end{figure}

By exploring a few candidate KAM surfaces, we found that the surface with $\iotabar=\varphi^{-2}= \lcontfrac 0,2,1,1,\cdots \rcontfrac \approx 0.382 $,
a noble number, is the last to break up between the major islands at $\iotabar = 0$ and $\iotabar = 1$.
The sequence $\lcontfrac a_0, a_1, a_2, \cdots \rcontfrac$ denotes the \textit{continued fraction expansion} of $\iotabar$, given by
\begin{equation}
    \iotabar = a_0 + \cfrac{1}{a_1 
            + \cfrac{1}{a_2 
             + \cdots } },
\label{eq:continued_fraction}
\end{equation}
with $a_0, a_1, \cdots \in \naturalnum^+$.
A noble number is an irrational number whose continued fraction expansion has all 1's on its tail.
Flux surfaces with a noble rotational transform are more resilient to chaos-inducing perturbations and more likely to persist~\cite{Greene1979}.
The critical value $\delta$ 
\modi
for this surface to break up 
\norm
is estimated to be $\delta_c =0.020782$ using the Greene's residue method~\cite{Greene1979} with the numerical package \textit{pyoculus}
\footnote{https://github.com/zhisong/pyoculus}.
An introduction to Greene's residue and the corresponding procedure to compute $\delta_c$ is given in \ref{app:greenes_residue}.
\modi
The \Poincare plot for $\delta = \delta_c$, zoomed into the region around this KAM surface is shown in \figref{fig:poincare_zoom}, with the arrow pointing towards the critical KAM surface.
Critical KAM surfaces are named as \textit{boundary circles} by Greene \cite{Greene1979},
which have chaos right next to them and will break up with an infinitesimal increase in the size of perturbation.
\norm
If $\delta < \delta_c$, the KAM surface with $\iotabar = \varphi^{-2}$ exists.
If $\delta > \delta_c$, this KAM surface doesn't exist.
If $\delta = \delta_c$, this KAM surface is a critical boundary circle.

\begin{figure}[htbp]
  \centering
  \[
    \begin{array}{c c}
      \includegraphics[width=8cm]{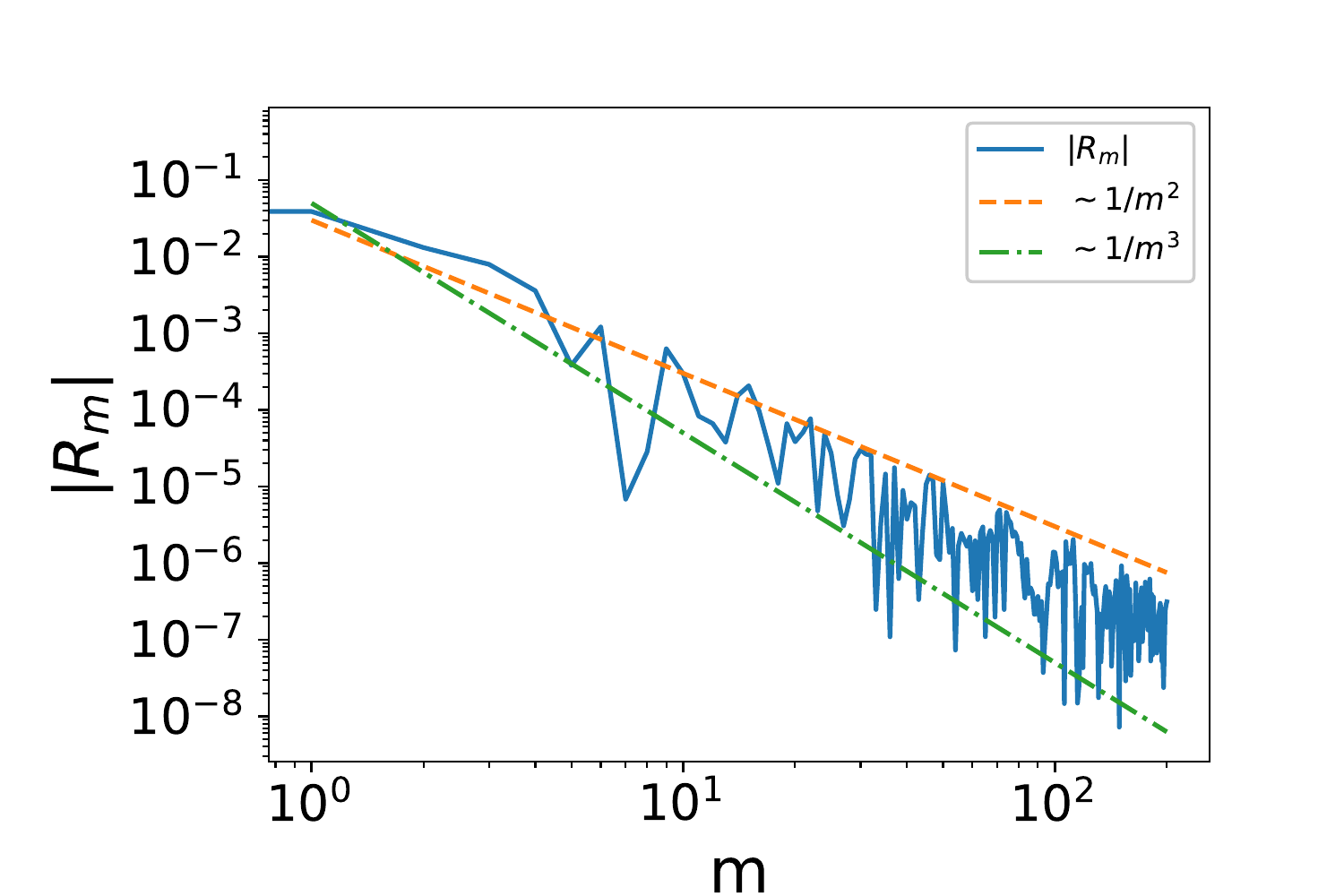} &
      \includegraphics[width=8cm]{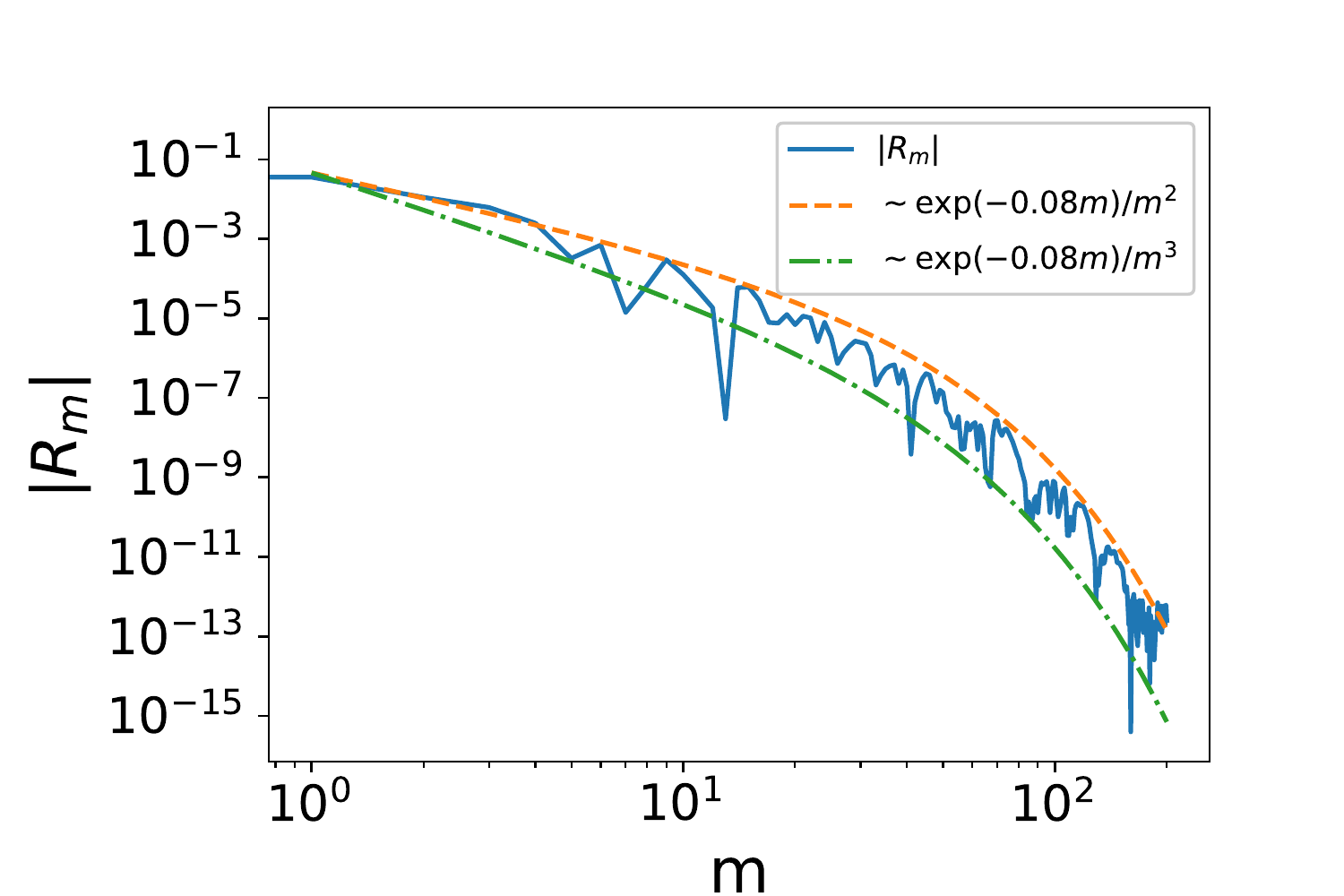} \\
      \text{(a) a boundary circle with } \delta=\delta_c  & \text{(b) } \delta=0.019<\delta_c
    \end{array}
    \]
  \caption{(a) The Fourier spectrum of the boundary circle (solid) with $\delta = \delta_c$ and the trend lines $\sim 1/m^2$ (dash) and $\sim 1/m^3$ (dash dot).
  (b) That of the sub-critical KAM surface with $\delta = 0.019 < \delta_c$ and the trend lines $\sim e^{-0.08m}/m^2$ (dash) and $\sim e^{-0.08m}/m^3$ (dash dot). }
  \label{fig:circle_Rm}
\end{figure}

A KAM surface can be represented by the functional form of a surface $R(\theta, \zeta)$.
To extract a KAM surface from the magnetic fields, Greene~\cite{Greene1979} suggested to approximate it by high order fixed points with rational $\iotabar$ converging to that of the KAM surface.
Using this method, we located the fixed points with $\iotabar = 610/1597$ for the case with $\delta = \delta_c$.
On the $\zeta =0$ plane, these fixed points form a set of discrete representation of $R(\theta, 0)$.
We fit the data with 200 Fourier harmonics and obtain its spectrum, as shown in \figref{fig:circle_Rm} (a).
It is worth stating that on a log-log scale, if one writes $R(\theta,0) = \sum_m R_m \cos m \theta$, the Fourier harmonics $|R_m|$ are roughly bounded between two lines,
representing a trend of $1/m^2$ and $1/m^3$, respectively.
This is slightly different from that of the standard map,
where the Fourier harmonics of a boundary circle are bounded between $1/m$ and $1/m^3$~\cite{Greene1981,Shenker1982}.
This does not converge to an analytic function.
\modi
On the other hand, the Fourier harmonics of a KAM surface below critical have an exponential converging rate.
We have also plotted the Fourier harmonics $\delta = 0.019 < \delta_c$ in \figref{fig:circle_Rm} (b), showing the exponential decay.
It means the Fourier series defining $R(\theta,0)$ converges on a strip around the real axis on the complex plane,
and as such, $R(\theta,0)$ is an analytic function.
\norm

\modi
The analyticity and smoothness properties of KAM surfaces need a further discussion.
Greene \cite{Greene1981} and Shenker and Kadanoff \cite{Shenker1982} showed in their numerical experiments that for the standard map the sub-critical KAM surfaces are analytic.
On the other hand, the boundary circles are continuous but their first derivatives are discontinuous.
In fact, these authors discovered that boundary circles have self-similar structures at all length scales, i.e. they are fractal surfaces.
In our system, we have also found the same for our KAM surface with $\iotabar = \varphi^{-2}$.
The detail is given in \ref{app:KAM_smoothness}.
\norm


\subsection{A three-volume HKT equilibrium}
\label{sec:three_volume}
\begin{figure}[htbp]
  \centering
  \[
    \begin{array}{c c}
      \includegraphics[width=8cm]{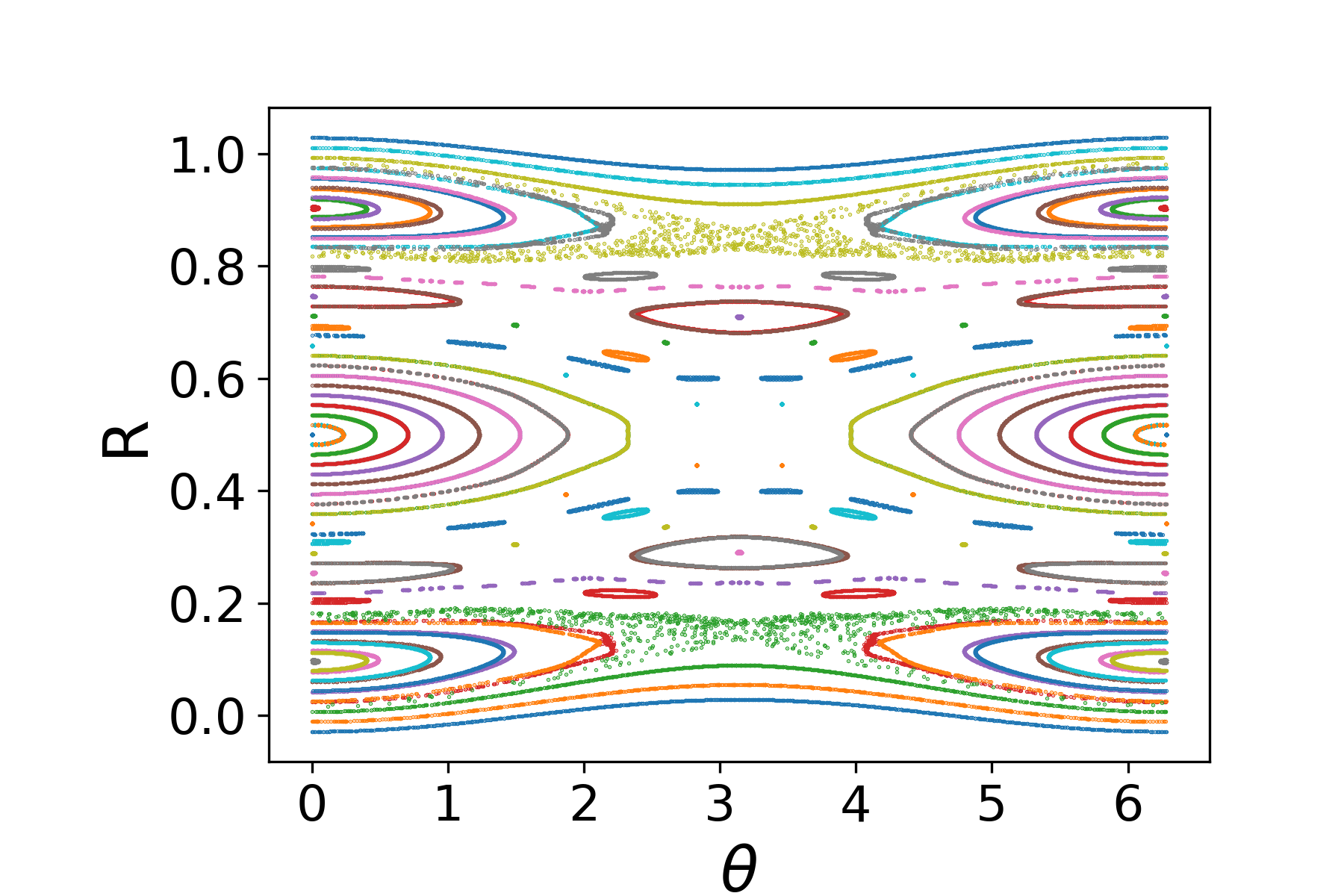} &
      \includegraphics[width=8cm]{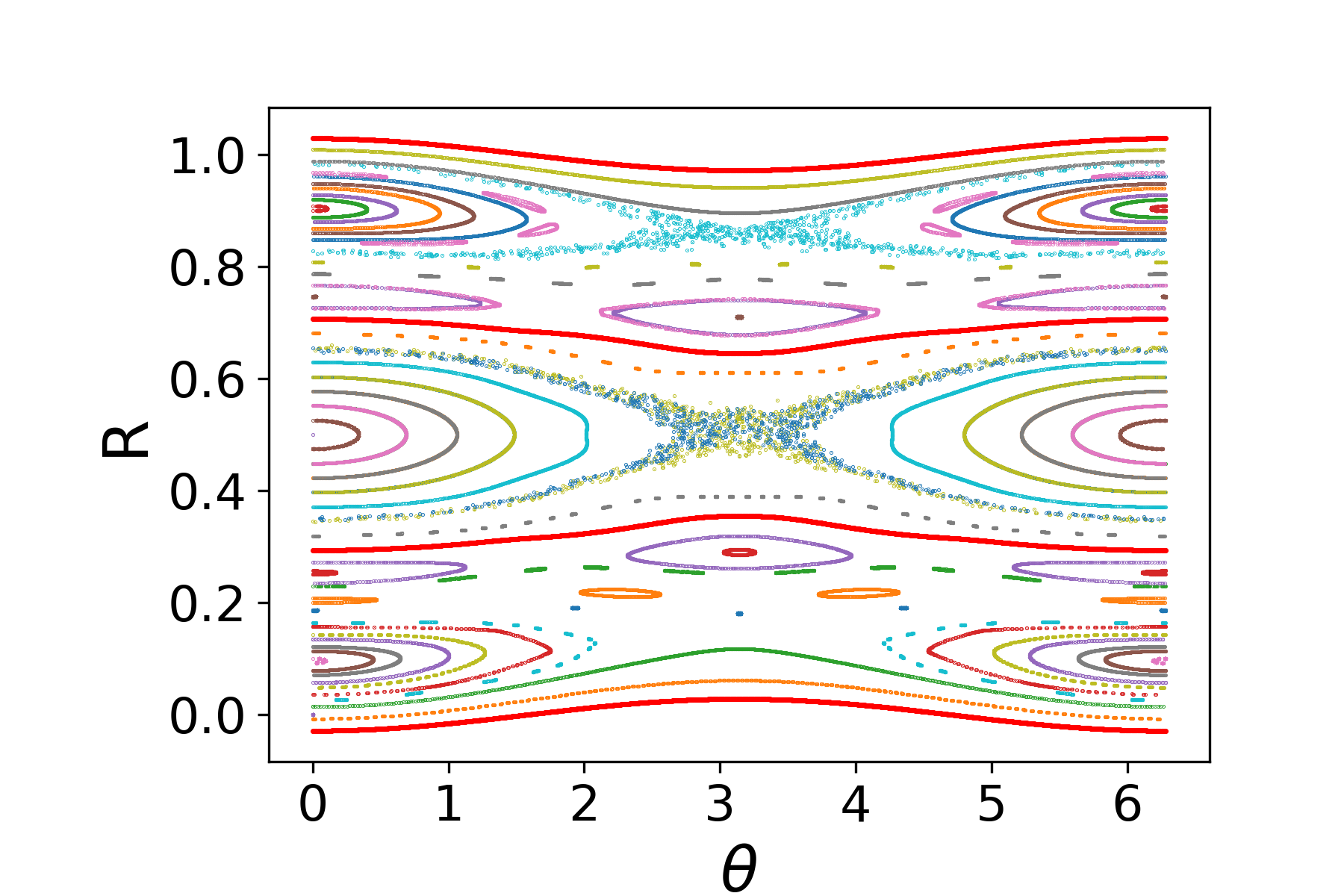} \\
      \text{(a)}  & \text{(b)}
    \end{array}
    \]
  \caption{The \Poincare plot of (a) one volume (b) three-volume HKT slab with $\delta=0.019 < \delta_c$.
  The red solid lines indicate the ideal interfaces and the plasma boundary.}
  \label{fig:poincare_plot_3vol}
\end{figure}

Let us first consider the cases $\delta < \delta_c$ when the KAM surface with $\iotabar = \varphi^{-2}$ still exists.
If we put an interface on this KAM surface, we should exactly reproduce the same one-volume equilibrium.
Since the upper and lower plasma boundary mirrors one another, instead of constructing a two-volume equilibrium,
we can put an additional interface at $\iotabar = -\varphi^{-2}$ which exactly mirrors the $\iotabar = \varphi^{-2}$ interface.
This leads to a three-volume equilibrium.
As mentioned earlier, a SPEC equilibrium needs three constraints for each volume.
We choose to constrain $\iotabar$ on the two bounding interfaces of the volume including on the plasma boundary,
taking up two constraints for each volume.
The last constraint, the toroidal flux $\psi_{t,i}$ in each volume $i$, remains unknown.
Due to the symmetry of the perturbed plasma upper/lower boundary, we choose for the first and third volume that $\psi_{t,1} = \psi_{t,3}$ and thus $\psi_{t,2}= 1-2\psi_{t,1}$ for the second volume,
such that $\psi_{t,1}+\psi_{t,2}+\psi_{t,3}=1$.
The total toroidal flux is normalized to unity, same as the one-volume equilibrium.
Therefore, the only undecided parameter is $\psi_{t,1}$.
Giving an initial guess of $\psi_{t,1}$, we iterate it to make $\mu$ the same in all three volumes.
By doing so, we expect the three-volume equilibrium to reproduce exactly the one-volume equilibrium, i.e.
the two added interfaces would just be the two KAM surfaces in the one-volume case, 
\modi
given $\mu$ the same across all volumes.
\norm
A demonstration of such a three-volume equilibrium for $\delta = 0.019$ is plotted in \figref{fig:poincare_plot_3vol} (b), with the corresponding one-volume equilibrium in \figref{fig:poincare_plot_3vol} (a).

\modi
It is worthy to mention that the uniqueness of multi-region stepped equilibria is not guaranteed.
In this paper, we have chosen the initial guess of the equilibria specifically to reproduce the one-volume HKT problem.
What we mean by ``existence of solutions'' is actually ``existence of solutions of a particular class''.
We are interested to study when a certain class of equilbria exists or ceases to exist.
This does not rule out the existence of another class of solution that is far from the single-volume HKT solution in \secref{sec:one_volume}.
\norm

The same process can be extended to the cases with $\delta > \delta_c$ when the $\iotabar = \varphi^{-2}$ surface is broken into cantori.
Here we choose $\delta = 0.022$ as an example. 
The solution, if it exists, has no equivalent single-volume equilibrium.
Similar to shielding an island by forcing a flux surface at the position of the island, 
the above solution is equivalent to forcing an interface on the cantori after the KAM surface is broken.
SPEC can indeed find a three-volume equilibrium with the force balance satisfied.
\modi
A question remains whether it is a valid solution.
\norm

\begin{figure}[htbp]
  \centering
  \includegraphics[width=8cm]{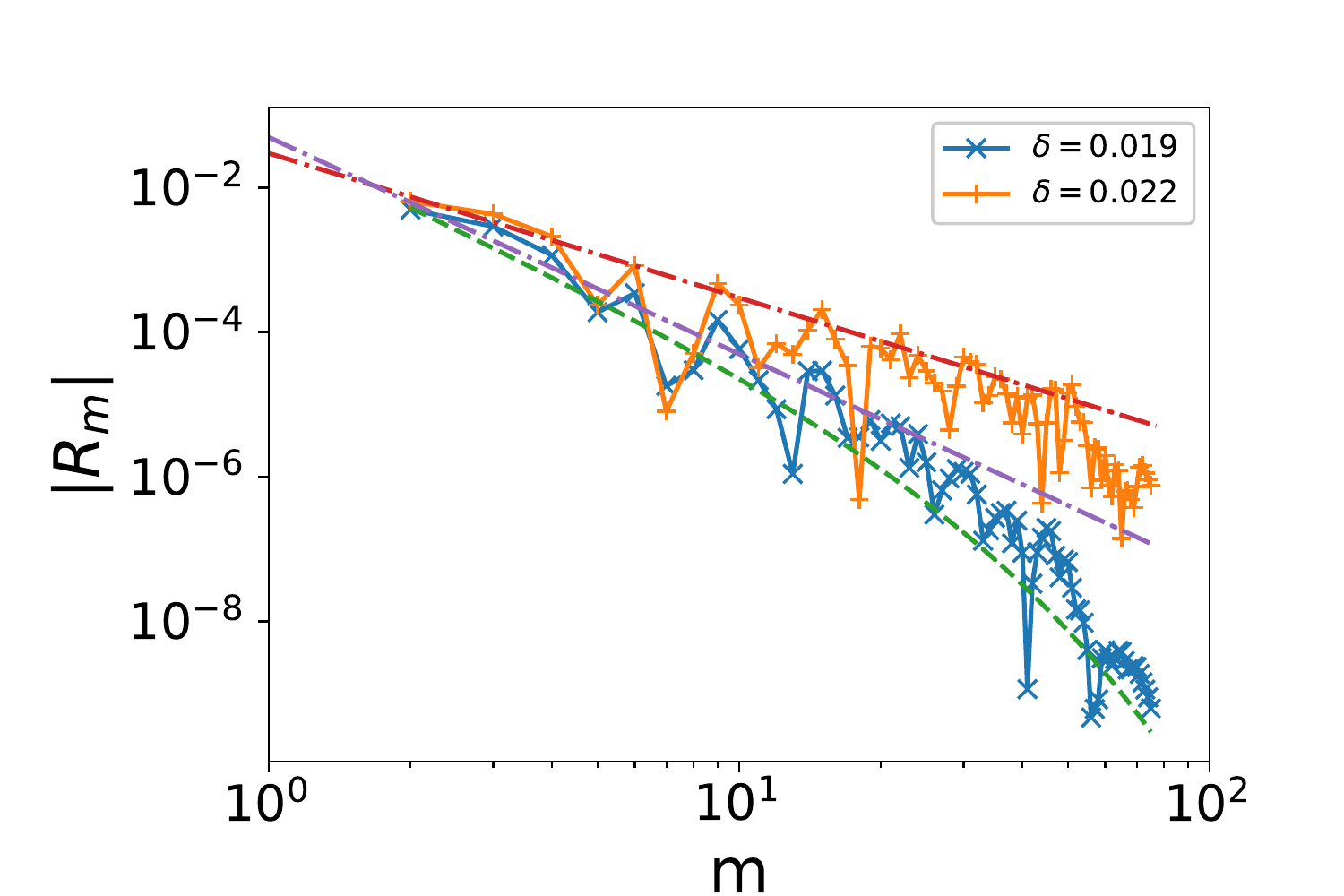}
  \caption{\modi
  The Fourier harmonics $|R_m|$ of $R(\theta, 0)$, the interface between the second and third volume in the three-volume HKT equilibrium,
  for $\delta = 0.019 < \delta_c$ and $\delta = 0.022 > \delta_c$ (solid lines) computed with $M=77$, $N=30$, and $L=10$.
  The harmonics are plotted in a log-log scale with $m$ being the poloidal Fourier mode number.
  The dashed lines \modii show \norm the trend.
  A straight line with slope $-2$ (upper dashed line) and $-3$ (middle dashed line) means a $1/m^2$ and $1/m^3$ trend, respectively, while a line curved downwards (lower dashed line) means an exponentially decaying trend.
  \norm}
  \label{fig:three_vol_Rm}
\end{figure}

\modi
In \figref{fig:three_vol_Rm} we show the Fourier harmonics $R_m$ for the $\delta =0.19 <\delta_c$ (under threshold) case
and the $\delta = 0.022 > \delta_c$ (over threshold) case on a log-log scale.
This SPEC equilibrium is computed with $M=77$, $N=30$, and $L=10$,
in which $M$ and $N$ are the poloidal and toroidal Fourier resolution, respectively,
and $L$ is the radial Chebyshev resolution.
We find that $R_m$ for the under threshold case decays exponentially as $m$ increases,
proving the interface to be an analytic surface.
On the contrary, $R_m$ for the over threshold case is bounded by two straight lines with the slope being $-2$ and $-3$, respectively,
matching that of a boundary circle in \figref{fig:circle_Rm} (a).
\modii
Therefore, the interface for $\delta  > \delta_c$ is fractal.
The solution with the requirement that the interface being a $C^1$ surface does not exist even in the weak sense,
since the fractal surface is everywhere non-$C^1$.
\norm

\modii
The Fourier based method in SPEC requires the interfaces to be analytic.
But when looking for a solution using an iteration procedure, the interface may well want to break the analyticity while remaining analytic at each iteration step.
\norm
An analogy is to minimise the function $f(x) = x^2$ within the range $x \in (0, +\infty)$.
A solution to the minimisation problem does not exist.
An optimizer will try to push closer towards $x=0$ which retaining $x>0$, but is never allowed to reach $x=0$.

\begin{figure}[htbp]
  \centering
  \[
    \begin{array}{c c}
      \includegraphics[width=8cm]{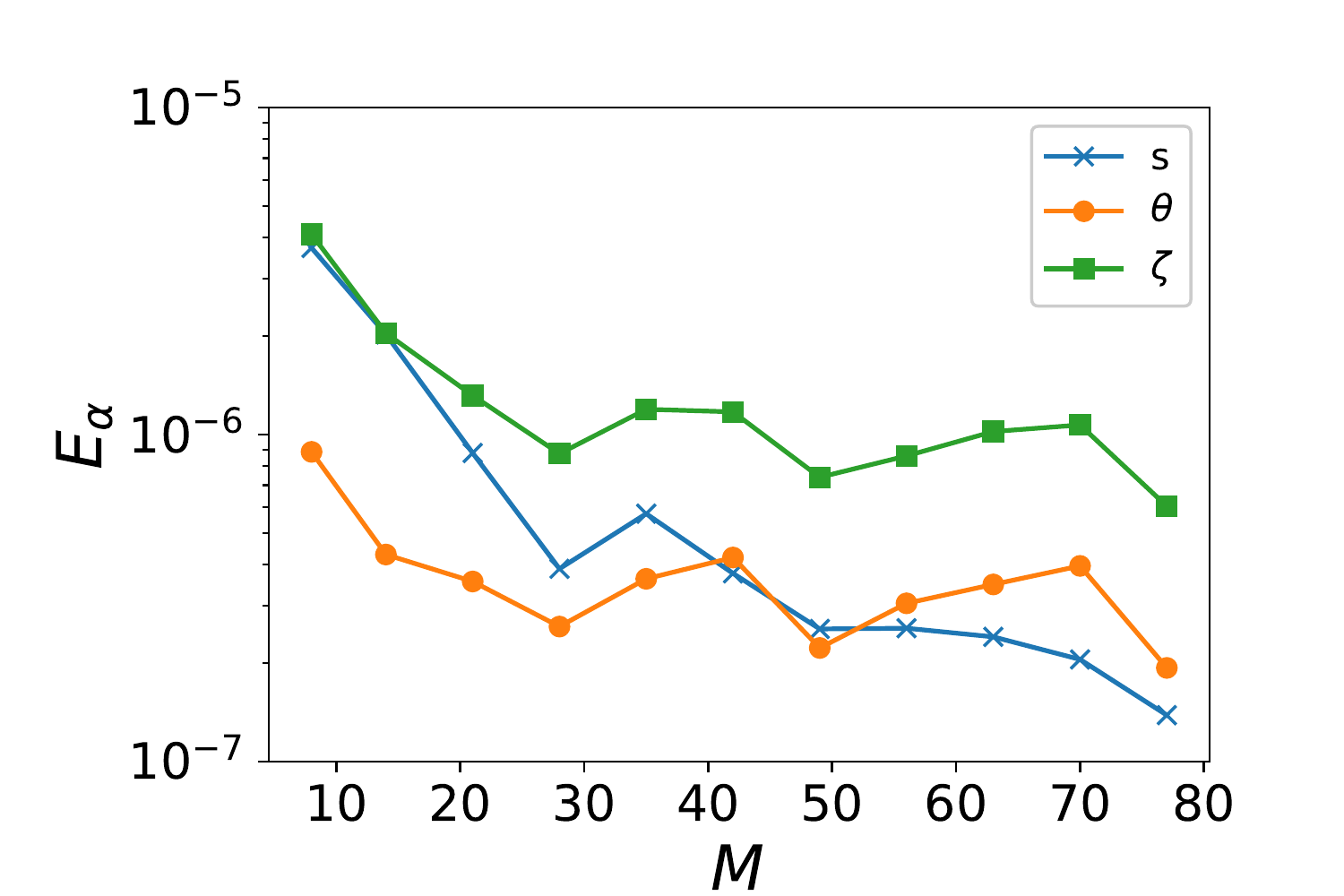} & 
      \includegraphics[width=8cm]{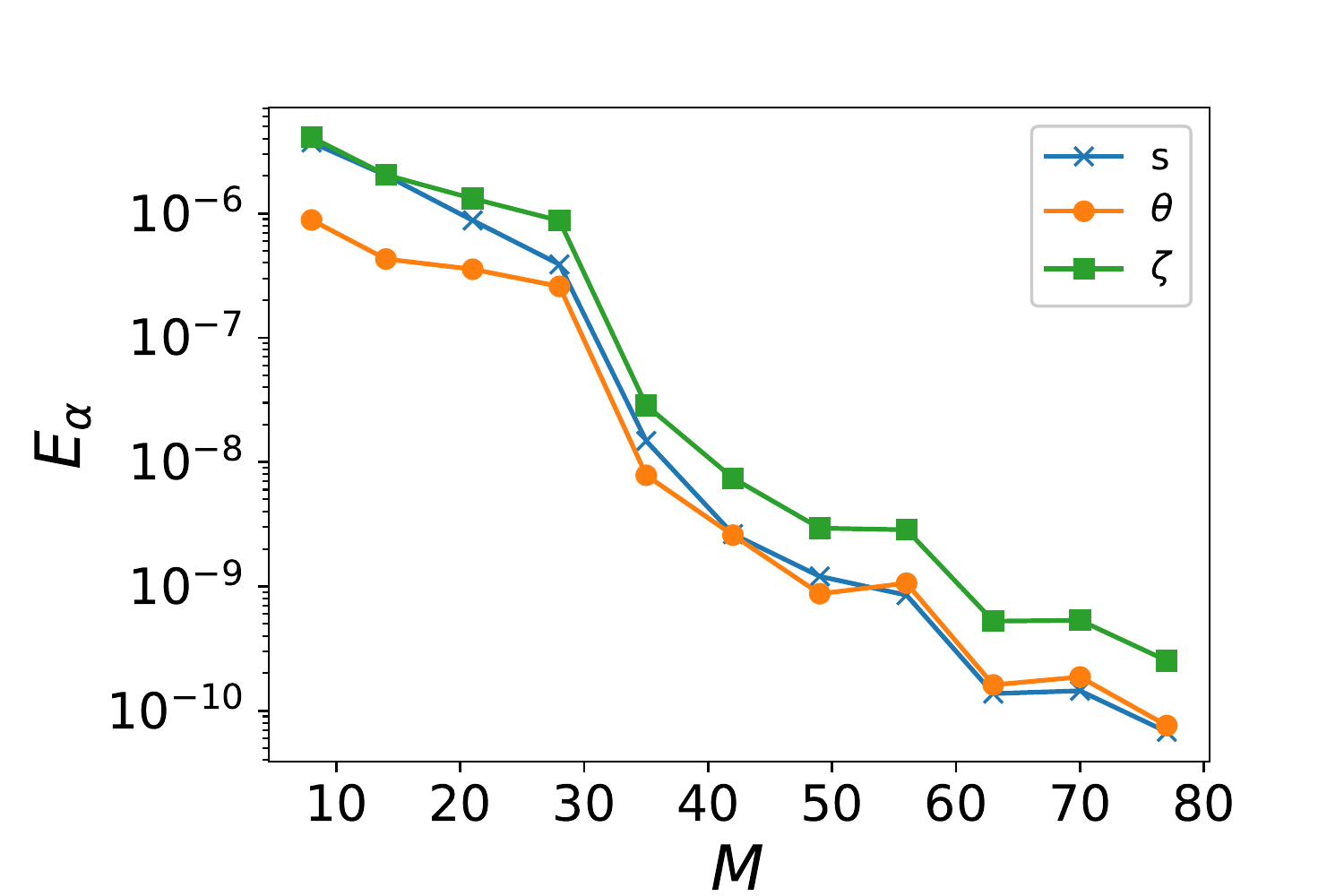} \\
      \text{(a)}  & \text{(b)}
    \end{array}
    \]
  \caption{
    The averaged error $E_\alpha$ for the middle volume in the (a) $\delta =0.022$ case and (b) $\delta =0.019$  as a function of poloidal Fourier resolution $M$, 
    where $\alpha = \{s, \theta, \zeta\}$ is the coordinate of interest. A straight line indicates exponential convergence.
  }
  \label{fig:beltrami_error}
\end{figure}

By examining the numerical error of the Beltrami solution, one can conclude that the SPEC solution to the $\delta = 0.022$ case does not converge even though it reaches force balance. 
\Figref{fig:beltrami_error}(a) shows the volume averaged error in the Beltrami equation defined as 
\begin{equation}
    E_\alpha = \frac{1}{V}\int_\Omega |(\nabla \times \bB - \mu \bB) \cdot \nabla \alpha| dV, 
\end{equation}
where $\alpha = \{s, \theta, \zeta\}$ is the coordinate of interest and $V$ the volume size.
A well converged SPEC solution has $E_\alpha$ exponentially decaying as the Fourier resolution increases, indicated by a straight line trending down on a log-linear scale~\cite{Loizu2016a}.
It is evident that the Beltrami error here saturates as the Fourier resolution increases, meaning that SPEC is not giving a converged solution.
This is as expected since the Fourier method used in SPEC assumes
\modii the interfaces as well as the magnetic fields to be
analytic and the method will have difficulties to converge otherwise.
\norm

As a comparison, \figref{fig:beltrami_error}(b) demonstrates the same error for the $\delta = 0.019$ case, showing clear exponential convergence.
\modii
However, our work strongly suggests that the convergence problem here is independent of the choice of numerical method but is rather due to the non-existence itself.
A grid-based method such as the finite element method may also struggle to converge.
\norm

\norm

In summary, a stepped-pressure equilibrium with given constraints and boundaries does not always exist.
One could be forcing an irrational interface on the cantori after the flux surface is broken.
If that happens, the interface Fourier harmonics do not converge exponentially.
This indicates one should consider removing that interface and merging the neighbouring two volumes. 

\section{Non-existence due to a finite pressure jump}
\label{sec:existence_PJH}

A non-symmetric interface can only support a finite pressure.
By utilizing the PJH framework, McGann~\cite{McGann2013} found that the KAM surface in the PJH phase space breaks up
for a large pressure jump on the interface.
If this happens, a solution to the magnetic field consistent with the force balance condition \eqref{eq:force_balance},
the surface current condition \eqref{eq:current_boundary_condition} and the given $\iotabar$, ceases to exist.
McGann discovered that a higher perturbation on the interface leads to a lower maximum allowed pressure jump.
In this section, we will first repeat McGann's method for our three-volume HKT problem in \secref{sec:three_volume}.
We then propose a new method based on the convergence property of the PJH return map,
which can give a sharper estimation than that of McGann.

\subsection{Allowed pressure jump estimated from Greene's residue}
\label{sec:PJH_McGann}
To estimate the maximum pressure jump, McGann took a SPEC interface,
kept it fixed, and scanned over $\Delta p$ until a solution no longer exists.
Similarly, we start from the pressure-less three-volume equilibria with different $\delta$ described by \secref{sec:three_volume}.
Now, we set the pressure in the first volume to be $p_1$
and that in the second volume to be $p_2$.
The third volume is set to mirror that in the first volume, with the same $p_3=p_1$.
The pressure jump at the second interface is $p_2 - p_1$  = $\Delta p$.
We will focus on the regime with $\Delta p \ge 0$, while the cases with $\Delta p \le 0$ can be similarly studied.
Note that the interfaces are left as they are in the pressure-less case.
\modi
In McGann's case the magnetic field in the second volume, while satisfying the force balance condition, 
is no longer a Beltrami field, or it is a Beltrami field but does not satisfy the boundary condition on the other boundary.
\norm
By doing so, we can isolate two different effects and study them separately: (1) the change in $\Delta p$, and (2) the change of interface geometry. 

\begin{figure}[htbp]
  \centering
  \[
    \begin{array}{c c}
      \includegraphics[width=8cm]{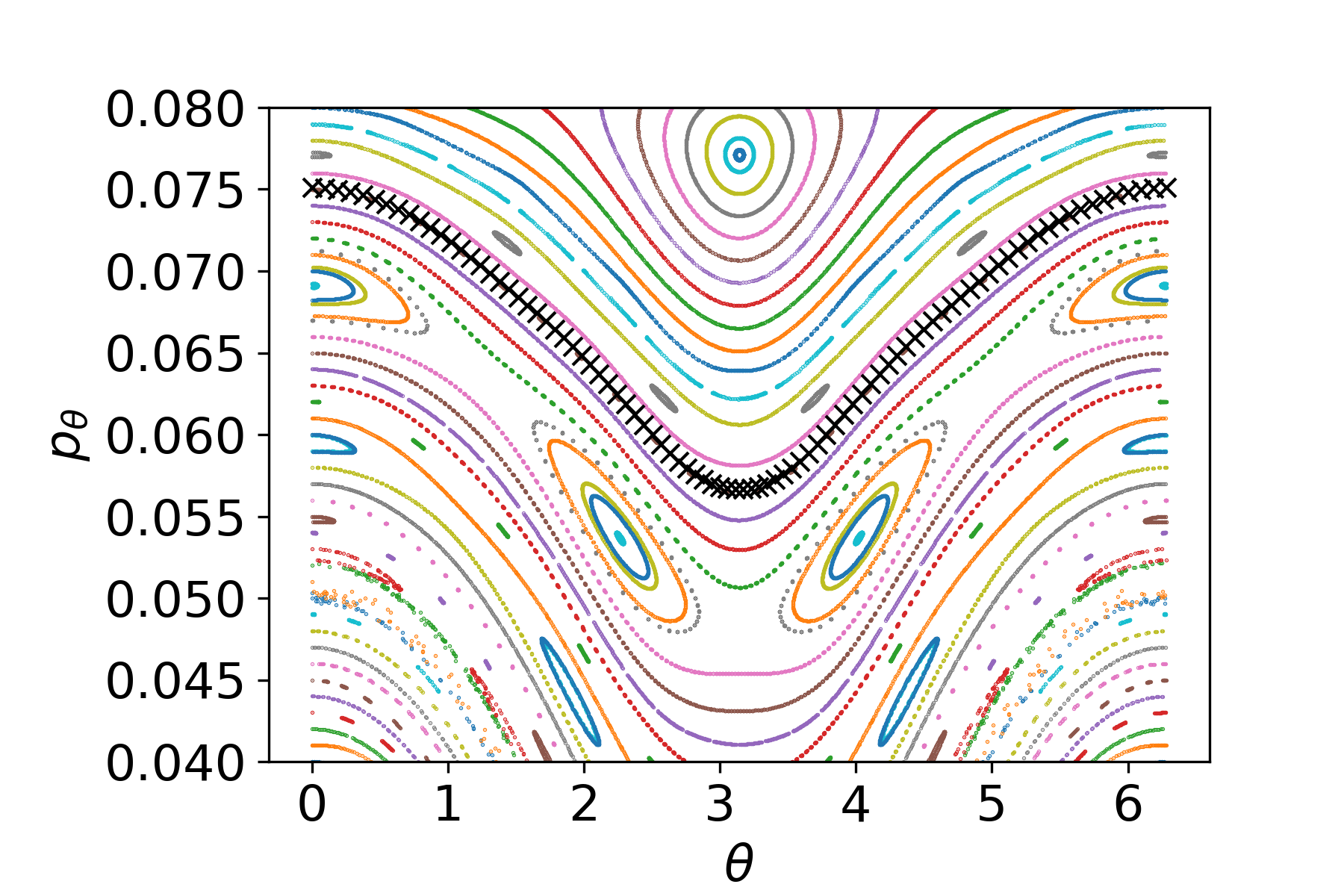} & 
      \includegraphics[width=8cm]{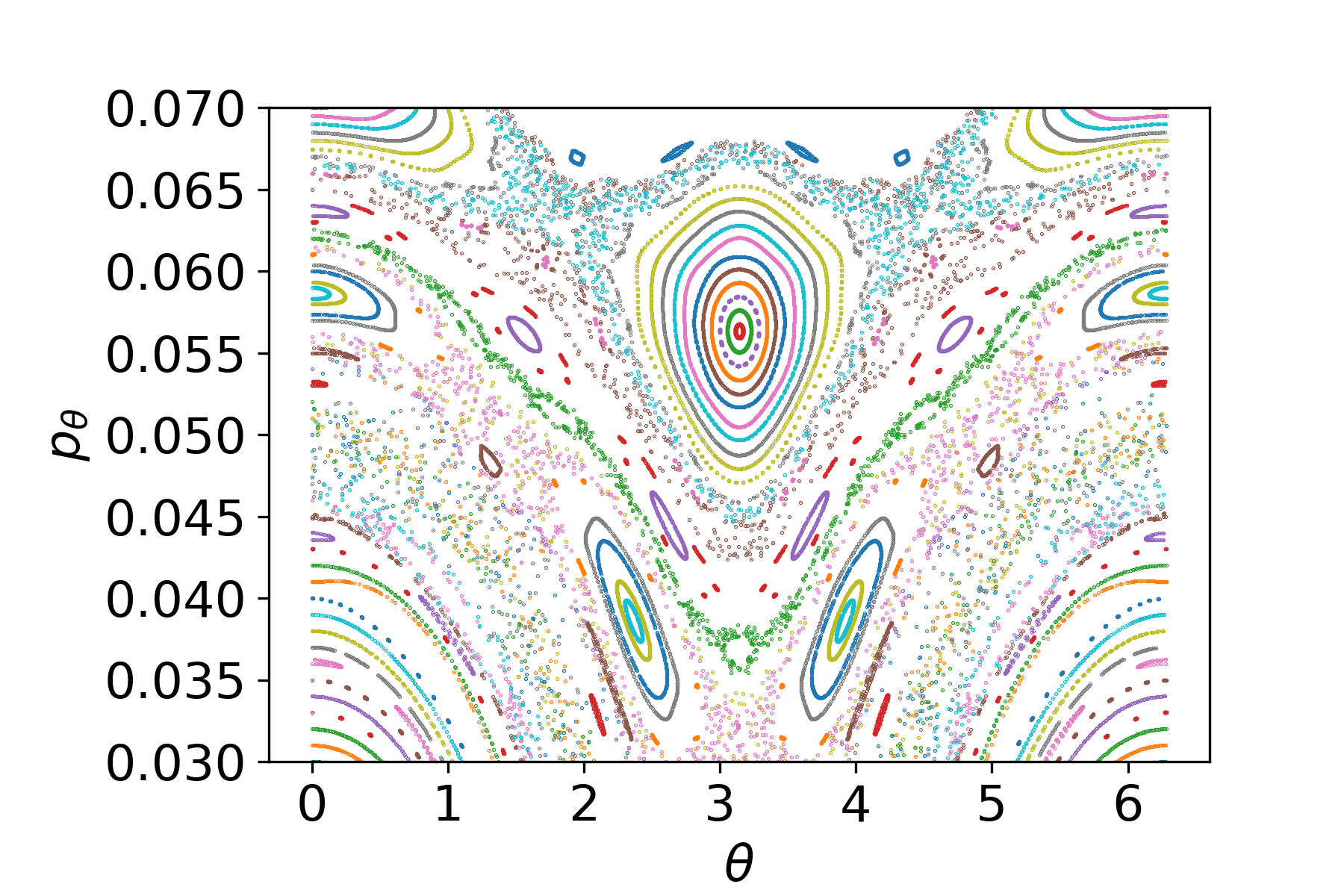} \\
      \text{(a)} & \text{(b)}
    \end{array}
  \]
  \caption{\Poincare plot of the PJH phase space, for the first interface with $\delta = 0.014$ and different $\Delta p$:
   (a) $\Delta p = 0.00078$ and (b) $\Delta p = 0.0078$.
  The KAM surface with $\iotabar=\varphi^{-2}$ is indicated by `x' markers, approximated by higher order fixed points of its convergent.
  The SPEC equilibria are computed with a resolution of $M=20, N=8$ and $L=10$.}
  \label{fig:PJH_Poincare_McGann}
\end{figure}

As an example, we generated the \Poincare plot in the phase space of the PJH for the first interface with $\delta = 0.014$,
by solving the Hamilton's 
\modi
equations \eqref{eq:PJH_dtheta} and \eqref{eq:PJH_dptheta} 
\norm
with the initial conditions $\theta=\zeta=0$ and a range of different $p_\theta$,
shown in \figref{fig:PJH_Poincare_McGann}.
The \Poincare plot of the PJH phase space looks much like that of the magnetic field,
but with the radial label $R$ replaced by $p_\theta$.
As in \secref{sec:preliminaries_PJH}, a solution to the force balance exists if and only if the corresponding KAM surface in the PJH phase space exists.
In \figref{fig:PJH_Poincare_McGann} (a) with $\Delta p = 0.00078$ (equivalent to a $\beta$ of $4\%$),
a KAM surface with $\iotabar = \varphi^{-2}$ exists, corresponding to a solution of the magnetic field in real space for the same $\iotabar$.
As $\Delta p$ increases, the phase space of the PJH become more chaotic.
For $\Delta p = 0.0078$ (a $\beta$ of $40\%$) as shown in \figref{fig:PJH_Poincare_McGann} (b), 
the KAM surface is already broken by the neighbouring chaos and a solution of the magnetic field in real space does not exist.

\begin{figure}[htbp]
  \centering
  \includegraphics[width=8cm]{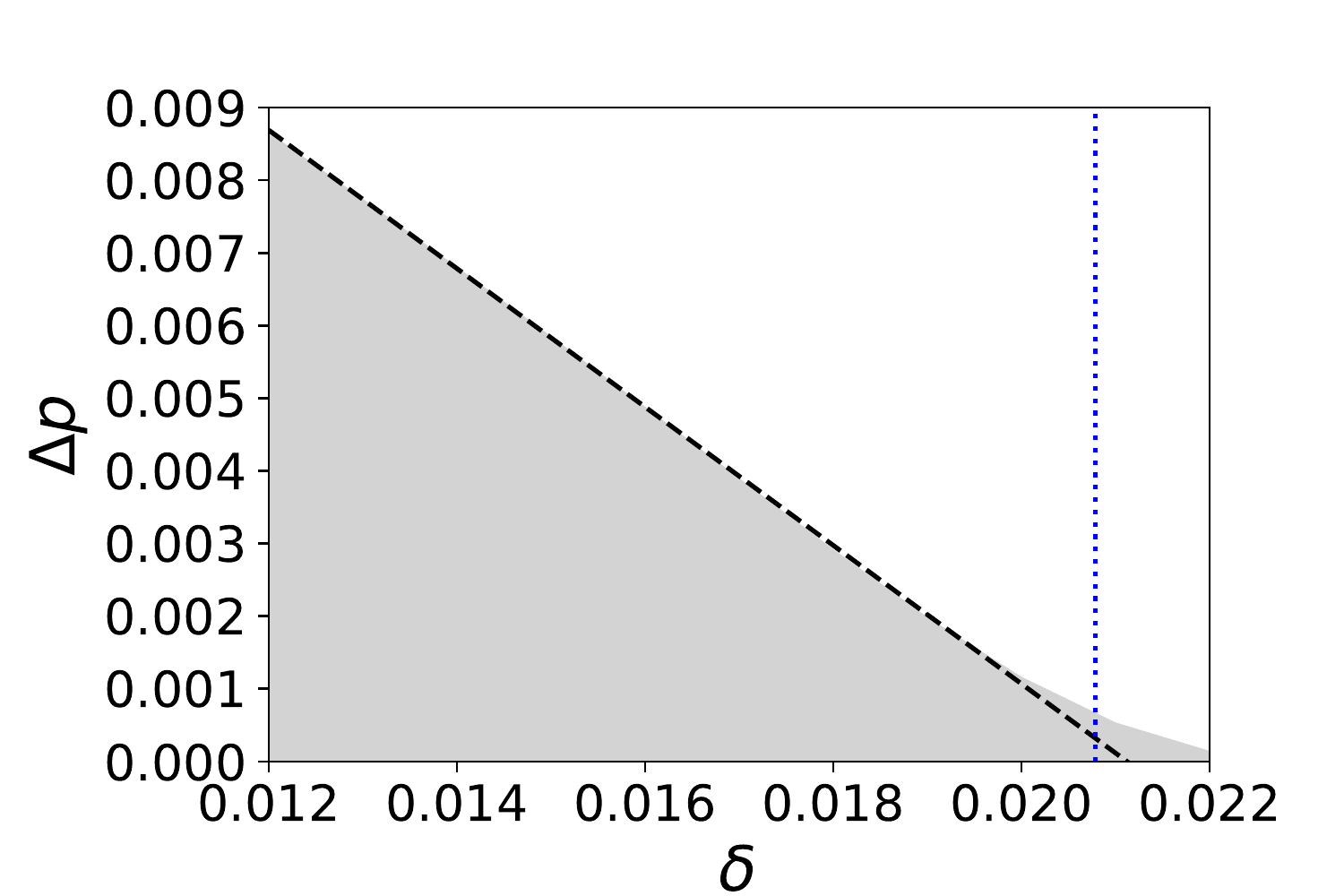}
  \caption{Allowed pressure jump (grey area) as a function of boundary perturbation $\delta$,
  estimated by Greene's residue applied to the PJH phase space.
  A linear fit of the boundary of the grey area using $\delta \le 0.18$ is marked by the dashed line.
  The vertical dotted line indicates the critical threshold $\delta_c=0.020782$.
  The SPEC equilibria are computed with a resolution of $M=20, N=8$ and $L=10$.
  }
  \label{fig:allowed_pressure_jump_McGann}
\end{figure}

The existence and criticality of the KAM surface can again be inferred from Greene's residue.
The next step is to iterate the value of $\Delta p$ such that the KAM surface in the PJH phase space is critical, and record the value of $\Delta p$.
This critical $\Delta p$ corresponds to the transition from the existence of a solution to the non-existence.
We repeat the same process for $0.012 \le \delta \le 0.022$ and plot the critical $\Delta p$ in \figref{fig:allowed_pressure_jump_McGann}.
The figure shows that the maximum allowed pressure jump on the interface is a linearly decreasing function of the size of the perturbation.
\modi
The reason for having a linear dependency is unknown at this stage, but may be of interest for future research.
\norm
Close to the interface break-up threshold, $\delta \approx \delta_c$, the critical $\Delta p$ curve becomes flattened.
Even when $\delta > \delta_c$, the interface appears to support a non-trivial pressure jump, but
this is a consequence of having an insufficient Fourier resolution in the SPEC equilibrium.
Note that all the equilibria here are computed with a SPEC resolution of $M=20, N=8$ and $L=10$ with the higher order harmonics truncated.
We found that using a higher resolution SPEC equilibrium and including more Fourier harmonics will bring the curve closer to the linear trend line,
but will not reduce the allowed pressure jump to zero for the $\delta > \delta_c$ cases:
this is a limitation of the current method.
If the curve was to follow the linear trend line, it will hit zero at a $\delta$ very close to $\delta_c$.
In summary, \figref{fig:allowed_pressure_jump_McGann} confirms that an interface closer to the break-up threshold can support less pressure jump.


McGann's method based on Greene's residue and a fixed interface can give a useful upper bound for the maximum pressure it can support,
as we will see in our later section.
It is simple and easy to implement, but has two major drawbacks.
First, once a new pressure jump is given, the equilibrium should be recomputed self-consistently to reach a new force balance.
Second, the truncation of the Fourier spectrum of an interface has a significant impact on the final result when the interface is close to break-up threshold.

\subsection{A non-existence criterion based on the convergence of the PJH return map}
\label{sec:zero_pressure_jump}
We propose to examine the convergence property of the PJH as the interface Fourier resolution increase,
instead of using the PJH directly in conjunction with Greene's residue criterion.
This will be the topic of the current section.

Although one can find out if an interface is smooth by examining the tail of its Fourier harmonics,
it is not always practical.
\modi
For example, to produce \figref{fig:three_vol_Rm}, we ran SPEC with 77 poloidal harmonics and 30 toroidal harmonics. 
The run takes two days with 48 cpus and 1.5TB memory in a three-volume slab and will be slower for a general toroidal case with more volumes.
\norm
Instead of running one case with an extremely high Fourier resolution, 
we can run the same case at a few lower Fourier resolutions and examine the convergence of the solution based on the PJH.
We will develop our method based on the series of three-volume equilibria with zero pressure jump in \secref{sec:kam_surface_break_up},
in which the non-existence of solution happens at $\delta > \delta_c = 0.020782$.

Picking an interface in the SPEC equilibrium, we can construct the corresponding PJH of that interface.
Taking $(B^-)^2$, the field on the inner side of the interface, to be the SPEC solution and prescribing the pressure jump $\Delta p$ to be same as the input parameters,
one can solve the PJH (locating the KAM surface with $\iotabar$) \modi by \norm finding the high order fixed points to obtain $B^+_\theta(\theta,\zeta)$ and $B^+_\zeta(\theta,\zeta)$,
the covariant magnetic field components on the outer side of the interface.
One can then compare them with those given by SPEC in the second volume, denoted by $B^{+'}_\theta(\theta,\zeta)$ and $B^{+'}_\zeta(\theta,\zeta)$,
and expect $|B^+_\theta - B^{+'}_\theta| = 0$ if all the equations \eqref{eq:Beltrami}, \eqref{eq:ideal_boundary_condition}
and \eqref{eq:force_balance} are exactly satisfied.
However, due to the finite numerical resolution, $|B^+_\theta - B^{+'}_\theta| \neq 0$.
There are several sources of numerical errors.
\begin{enumerate}
  \item The force balance condition \eqref{eq:force_balance} is solved by reducing its Fourier harmonics to zero. 
  The Fourier harmonics are truncated according to the Fourier resolution in a SPEC run. 
  \item There are numerical errors in solving the Beltrami equation \eqref{eq:Beltrami} due to discretisation.
  \Eqref{eq:current_boundary_condition} is not exactly satisfied in SPEC.
  \item There are numerical errors associated with locating the KAM surface in the PJH phase space.
\end{enumerate}
The errors (i) and (ii) should converge to zero as the numerical resolution of SPEC increases.
The spectral method in SPEC ensures both of them to decay exponentially as a function of the Fourier resolution,
providing the interface is analytic.
Once the KAM surface is computed to very high precision and error (iii) is eliminated,
one would expect $|B^+_\theta - B^{+'}_\theta|$ to converge to zero at an exponential rate.

To locate the KAM surface in the PJH phase space accurately and thereby minimize error (iii), 
one needs to find the periodic orbit corresponding to a high order convergent of the irrational $\iotabar$,
i.e. the orbit in phase space should be followed for a sufficiently long distance.
This leads to large numerical error accumulation in the ODE integrator such as the fourth order Runge-Kutta 
and thus deteriorates the accuracy.
To avoid this difficulty, we take an alternative approach by comparing the return map of the PJH to the magnetic field given by SPEC on the other side of the interface.

More precisely, let $(\theta_0, \zeta_0)$ be an arbitrary point on the interface and $B^{+'}_{\theta,0} = B^{+'}_\theta(\theta_0, \zeta_0)$
be the field on the outside of the interface given by SPEC.
We set the initial condition of a PJH orbit to be $p_\theta = p_{\theta,0} =B^{+'}_{\theta,0}$, $\theta = \theta_0$ and $\zeta = \zeta_0$.
If the PJH is consistent with the SPEC solution on the outer side of the interface,
this initial point should be on a KAM surface in the PJH phase space with the same rotational transform as the field lines in real space.
The next step is to integrate \eqref{eq:PJH_dtheta} and \eqref{eq:PJH_dptheta} until $\zeta = \zeta_0 + 2 \pi$ 
when the orbit penetrates the same plane $\zeta = \zeta_0$,
giving $(\theta_1, p_{\theta,1})$.
That is, we compute the return map of the orbit starting from $(\theta_0, p_{\theta,0})$ on the $\zeta = \zeta_0$ plane,
or in other words, the next point of its \Poincare section.
Again, if the PJH is consistent with the SPEC solution, $p_{\theta,1}$ should coincide with $B^{+'}_{\theta,1} = B^{+'}_\theta(\theta_1, \zeta_0)$,
the SPEC magnetic field in real space at the same point $(\theta_1, \zeta_0)$.

Due to numerical errors, $B^{+'}_{\theta,1} \neq p_{\theta,1}$,
but the difference should converge to zero as the Fourier resolution in SPEC increases.
We therefore define the relative difference, averaged on the interface, as a measure of convergence, written as
\begin{equation}
  e_{\text{avg}} = \frac{\langle |B^{+'}_{\theta,1}(\theta_0, \zeta_0) - p_{\theta,1}(\theta_0, \zeta_0)| \rangle}{\langle |B^{+'}_\theta| \rangle}
\end{equation}
in which $\langle \cdots \rangle$ denotes surface average and is in practice averaged over an equidistant grid in $\theta_0$ and $\zeta_0$.
The value of $e_{\text{avg}}$ is a function of the SPEC Fourier resolution,
which can be used to infer the existence of a solution.
Empirically, we find that if a solution exists, $e_{\text{avg}}$ has the relationship
\begin{align}
  e_{\text{avg}} \sim \frac{C}{M^3} \exp(\alpha M),
\end{align}
in which  $M$ is the SPEC poloidal Fourier resolution and $\alpha$ and $C$ are constants.
The most likely reason for this empirical relationship is that the Fourier harmonics of a boundary circle
have a lower trend line of $1/m^3$, as shown in \figref{fig:circle_Rm}.
We have chosen the toroidal Fourier resolution 
and the radial Chebyshev resolution such that they are not the bottleneck of the resolution.

\begin{figure}[htbp]
  \centering
  \includegraphics[width=8cm]{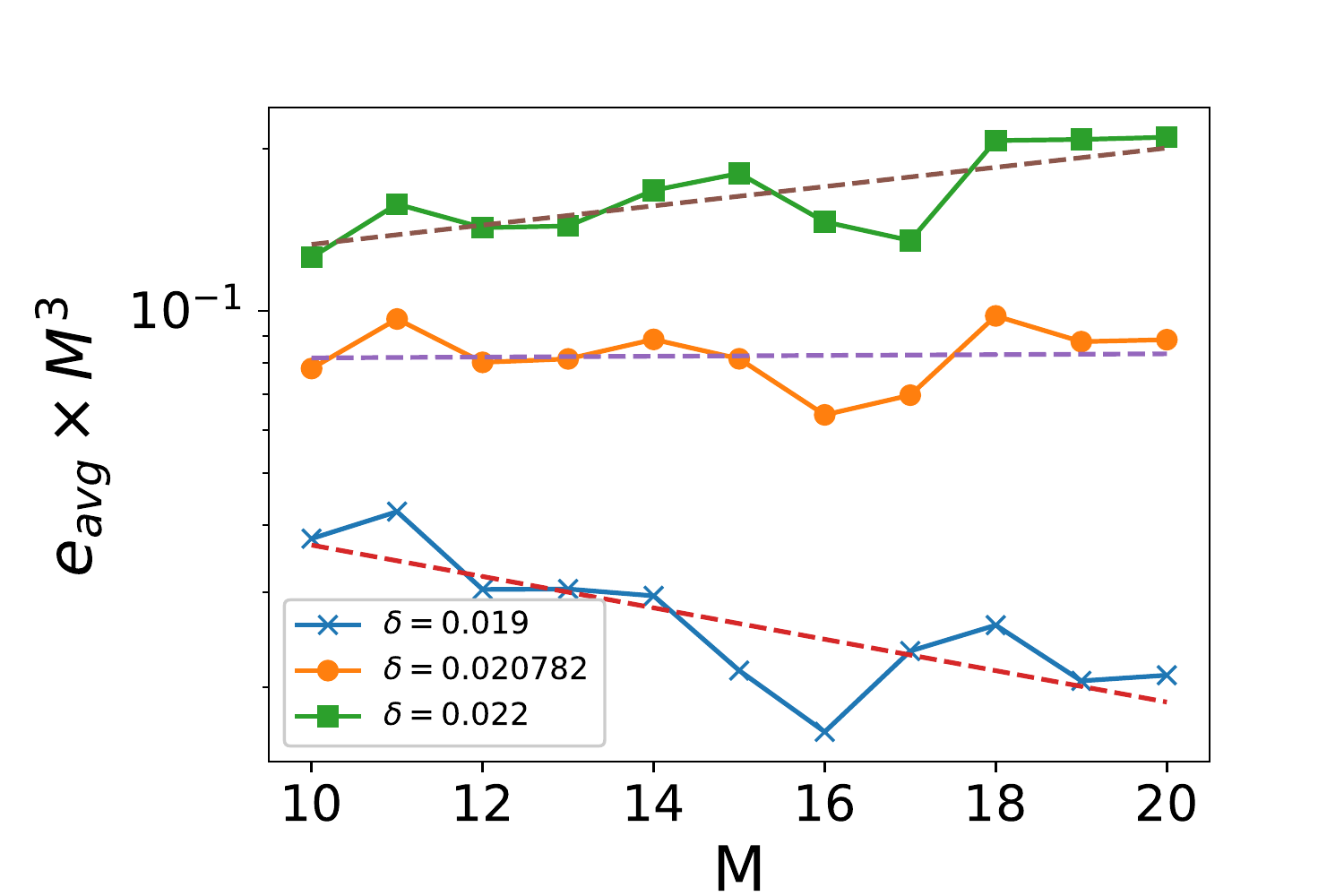}
  \caption{The averaged difference between the PJH return map and the magnetic field in real space,
  $e_{\text{avg}}$, multiplied by $M^3$, the SPEC poloidal Fourier resolution, plotted on a log-log scale,
  as a function of $M$, for sub-critical, critical and over-critical interfaces. 
  The dashed lines are linear least square fit of the curves.}
  \label{fig:eavg_no_pressure}
\end{figure}

\begin{figure}[htbp]
  \centering
  \includegraphics[width=8cm]{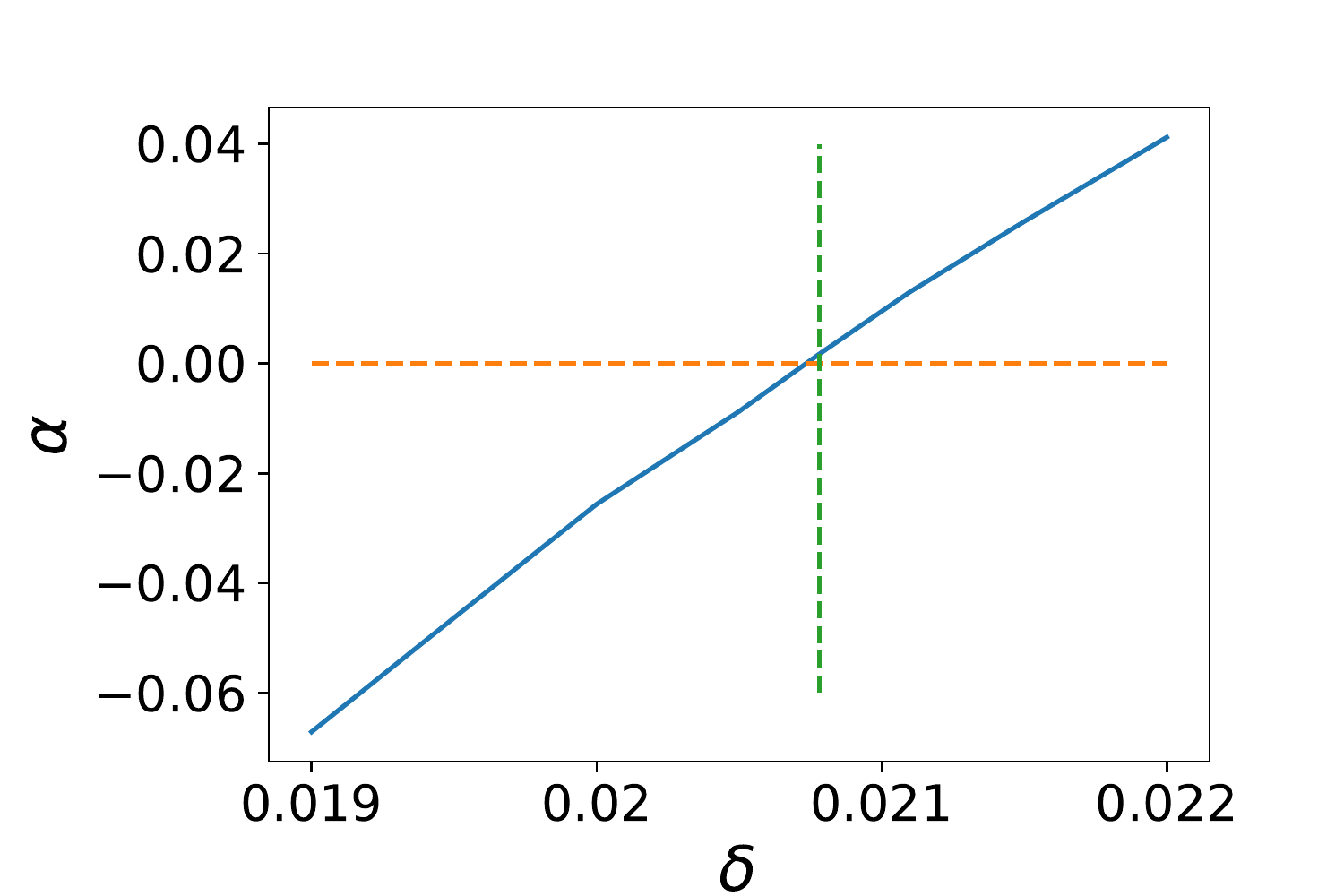}
  \caption{The fitting coefficient $\alpha$, the exponential decay factor of $e_{\text{avg}}$ versus $M$, as a function of $\delta$.
  The vertical line corresponds to the critical $\delta_c$ found by the Greene's residue criterion in the single-volume equilibrium.}
  \label{fig:alpha_delta_no_pressure}
\end{figure}

\Figref{fig:eavg_no_pressure} plots the quantity $e_{\text{avg}} M^3$ on a log-linear scale for $10\le M \le 20$,
for three cases $\delta =0.019 < \delta_c$ (under threshold), $\delta =0.020782 = \delta_c$ (critical)
and $\delta =0.022 > \delta_c$ (over threshold).
We have chosen this range of $M$ so it is large enough to show the exponential decay of $e_{\text{avg}}$ as a function of $M$, 
while small enough to be computationally practical.
For example, the scan over $10\le M \le 20$ can be completed typically within 30 minutes for each $\delta$.
Inspection of \figref{fig:eavg_no_pressure} shows that these curves lie around straight lines on a log-linear scale.
The coefficient of the exponential term, $\alpha$, 
is the slope of the straight line,
which can be computed by a linear least square fitting.
For the under threshold case, $\alpha < 0$ and $e_{\text{avg}}$ is dominated by the exponential decay for large $M$ when trending towards zero.
For the critical case, $\alpha = 0$ and there is no exponential decay.
The trend line is almost horizontal.
For the over threshold case, $\alpha > 0$, it is unclear if $e_{\text{avg}}$ will converge.
We have computed $\alpha$ using the same method for $0.019 \le \delta \le 0.022$ as shown in \figref{fig:alpha_delta_no_pressure}.
We found that the value of $\delta$ corresponding to $\alpha =0$ matches $\delta_c$ estimated by the Greene's residue criterion in the single-volume equilibrium.
This verifies that our method can give the correct value $\delta_c$ without the need to refer to the single-volume case.
If a different range of $M$ is chosen, the prediction of $\delta_c$ will be slightly different,
but is in general between $0.02$ and $0.021$, with the correct $\delta_c$ being $0.020782$.
Our current choice of $M$ range gives the best prediction.

The method proposed here has some advantages.
It does not require the equilibrium to have a single-volume counterpart and allows a finite pressure jump on the interface. 
The calculation is 
\modi
arguably
\norm
faster than running a single case with a very high Fourier resolution as it requires a much lower $M$. 
It also has some disadvantages.
First, the method is based on an empirical scaling law rather than 
\modi being 
mathematically rigorous.
\norm
Second, the calculation is truncated at a given Fourier resolution: 
it attempts to infer the convergence property at a higher Fourier resolution from the information at lower resolutions.
Finally, the applicability beyond the slab problem is untested at this stage.
\modi
Nevertheless, the method provides a way to identify cases which do not show a clear sign of convergence within the working range of Fourier resolution,
an indication that a solution with the desired analyticity property does not exist.
\norm

\subsection{Allowed pressure jump estimated from the PJH return map convergence}
\begin{figure}[htbp]
  \centering
  \includegraphics[width=8cm]{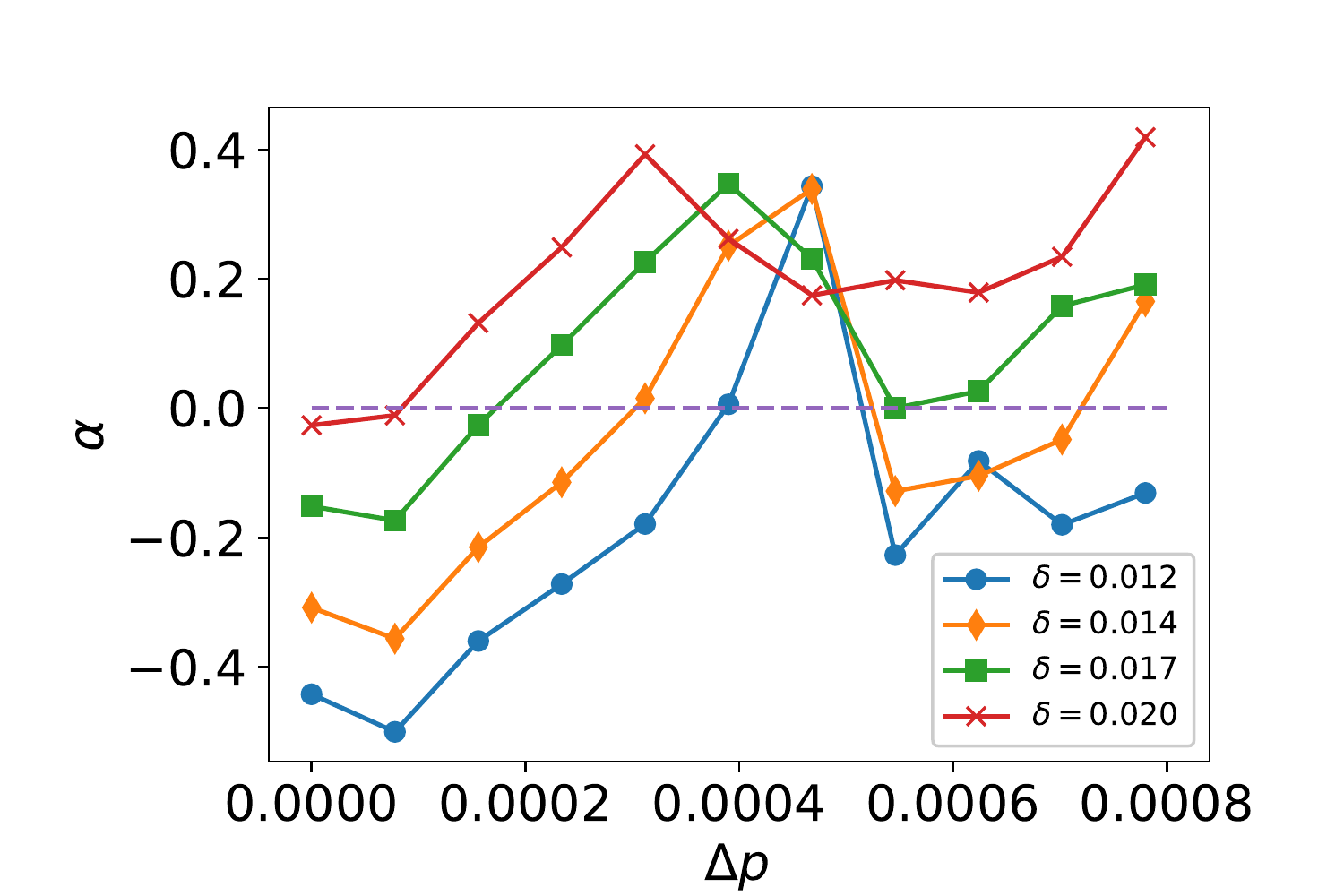}
  \caption{The exponential factor $\alpha$ as a function of interface pressure jump, for three choices of $\delta$.
  A positive $\alpha$ indicates the non-existence of solution with analytic interfaces.}
  \label{fig:alpha_pressure}
\end{figure}

Now we apply the method developed in \secref{sec:zero_pressure_jump} to the cases with a finite pressure jump.
We scan over the parameters space of $0.012 \le \delta \le 0.020782$ and $0 \le \Delta p \le 7.9\times 10^{-4}$
and construct the corresponding SPEC equilibria.
Note that $\Delta p =  7.9\times 10^{-4}$ corresponds to a pressure jump $\sim 4\%$ in $\beta$.
For each combination of $\delta$ and $\Delta p$, we obtain the exponential factor $\alpha$.
A plot of $\alpha$ versus $\Delta p$ for four different values of $\delta$ is shown in \figref{fig:alpha_pressure}.
\modi
A positive $\alpha$ indicates the non-existence of solutions with analytic interfaces.
Note that this is not yet equivalent to the non-existence of solutions,
but we will show the equivalency towards the end of the section.
\norm

\begin{figure}[htbp]
  \centering
  \includegraphics[width=8cm]{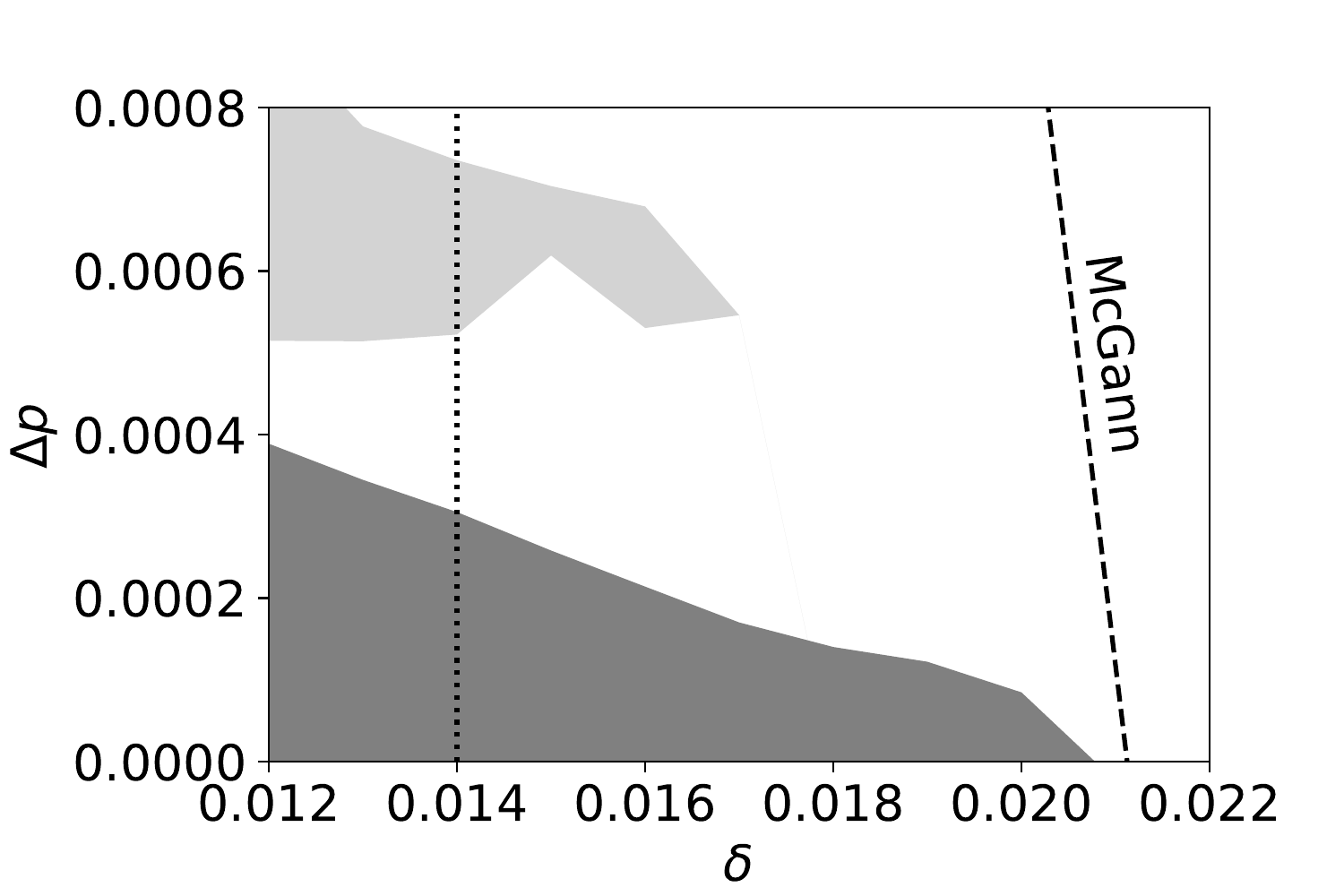}
  \caption{The $\Delta p -\delta$ parameter space for which a solution exists ($\alpha <0$, shaded) and does not exist ($\alpha > 0$, not shaded).
  The first and second existence regions are marked in grey and light grey, respectively.
  The vertical dotted line indicates the cases with $\delta =0.014$, which are further studied in subsequent figures.
  The black dash line shows the maximum pressure estimated by McGann's method shown in \figref{fig:allowed_pressure_jump_McGann}, i.e. taking the interface to be fixed while scanning $\Delta p$.
  }
  \label{fig:allowed_pressure}
\end{figure}

\Figref{fig:alpha_pressure} shows that $\alpha$ is not a monotonic function of $\Delta p$.
It first increases with $\Delta p$, peaks at around $\Delta p = 3.5 \times 10^{-4}$, and drops again.
The first non-existence region appears near the peak.
For a larger $\Delta p$ and small enough $\delta$, $\alpha$ drops to below zero.
This is consistent with the ``pressure healing'' effect discovered by McGann, 
which results in field configurations coming back into existence as the pressure discontinuity is made larger.
\modi
In other words, one cannot simply consider the pressure jump to play the same role as a symmetry-breaking perturbation in a near-integrable Hamiltonian system.
Rather, it is a different ``energy slice'' of the PJH.
\norm
The PJH is also significantly modified by the changed interface geometry due to force balance as $\Delta p$ changes.
After reaching its minimum, $\alpha$ monotonically increases to above zero, indicating a second non-existence region.
As $\delta$ increases, the whole $\Delta p - \alpha$ curve shifts upwards.
Finally, the first and second region of non-existence 
\modi
becomes
\norm
connected as $\delta$ further increases, giving a single $\Delta p$ threshold.
\Figref{fig:allowed_pressure} demonstrates the region in the $\Delta p - \delta$ parameter space for which $\alpha <0$ and thus the solution exists.
There are clearly two regions of existence and two regions of non-existence.
The two non-existence region are connected at $\delta \ge 0.017$.
As $\delta$ approaches $\delta_c=0.020782$, the allowed $\Delta p$ reduces to zero,
meaning that an interface at the threshold of breaking up cannot support any pressure.
We have overplotted the maximum pressure estimated using the method proposed by McGann as in \figref{fig:allowed_pressure_jump_McGann}.
It shows that the pressure threshold is much higher if the interface is taken to be fixed.
\modi
We have also scanned $\Delta p$ up to $0.002$ on a coarser grid, but we have not found another existence region,
although this does not rule out the possibility that for some $\Delta p > 0.002$ the equilibrium may come back into existence due to pressure healing.
\norm

\begin{figure}[htbp]
  \centering
  \includegraphics[width=8cm]{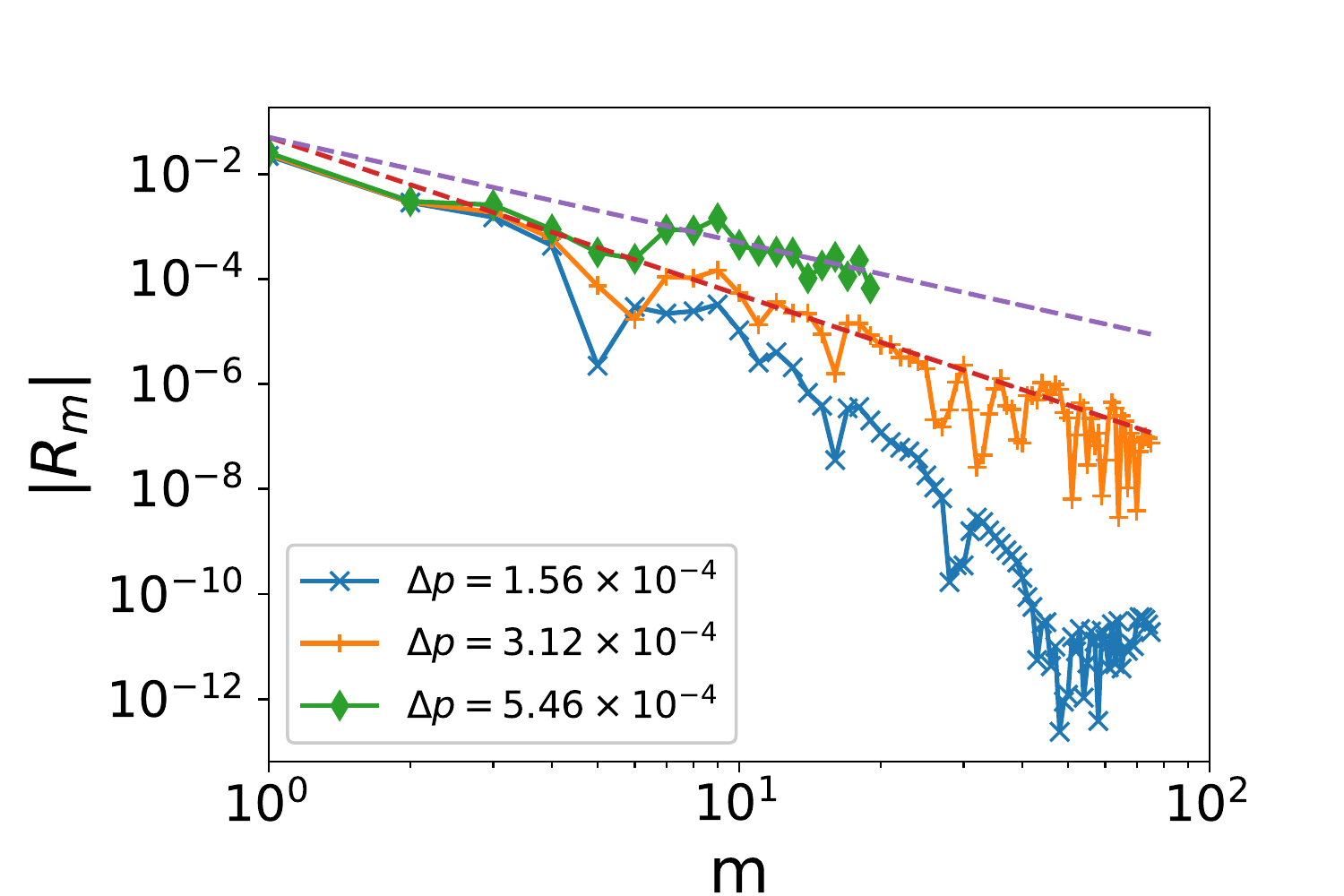}
  \caption{The Fourier harmonics $|R_m|$ of $R(\theta, 0)$,
  \modi of \norm
  the interface between the second and third volume the three-volume HKT equilibrium,
  for $\delta = 0.014$ and three different $\Delta p$ (solid lines), 
  \modi
  computed with $M=77$, $N=30$, and $L=10$. 
  The case with $\Delta p = 5.46 \times 10^{-4}$ is computed with $M=21$, $N=9$ and $L=10$.
  SPEC ceases to reach a force balanced solution for $M>22$.
  The harmonics are plotted in a log-log scale with $m$ being the poloidal Fourier mode number.
  The upper and lower dashed lines show trend lines of $m^{-2}$ and $m^{-3}$, respectively.
  \norm
  }
  \label{fig:Rm_delta014}
\end{figure}

\begin{figure}[htbp]
    \centering
    \includegraphics[width=8cm]{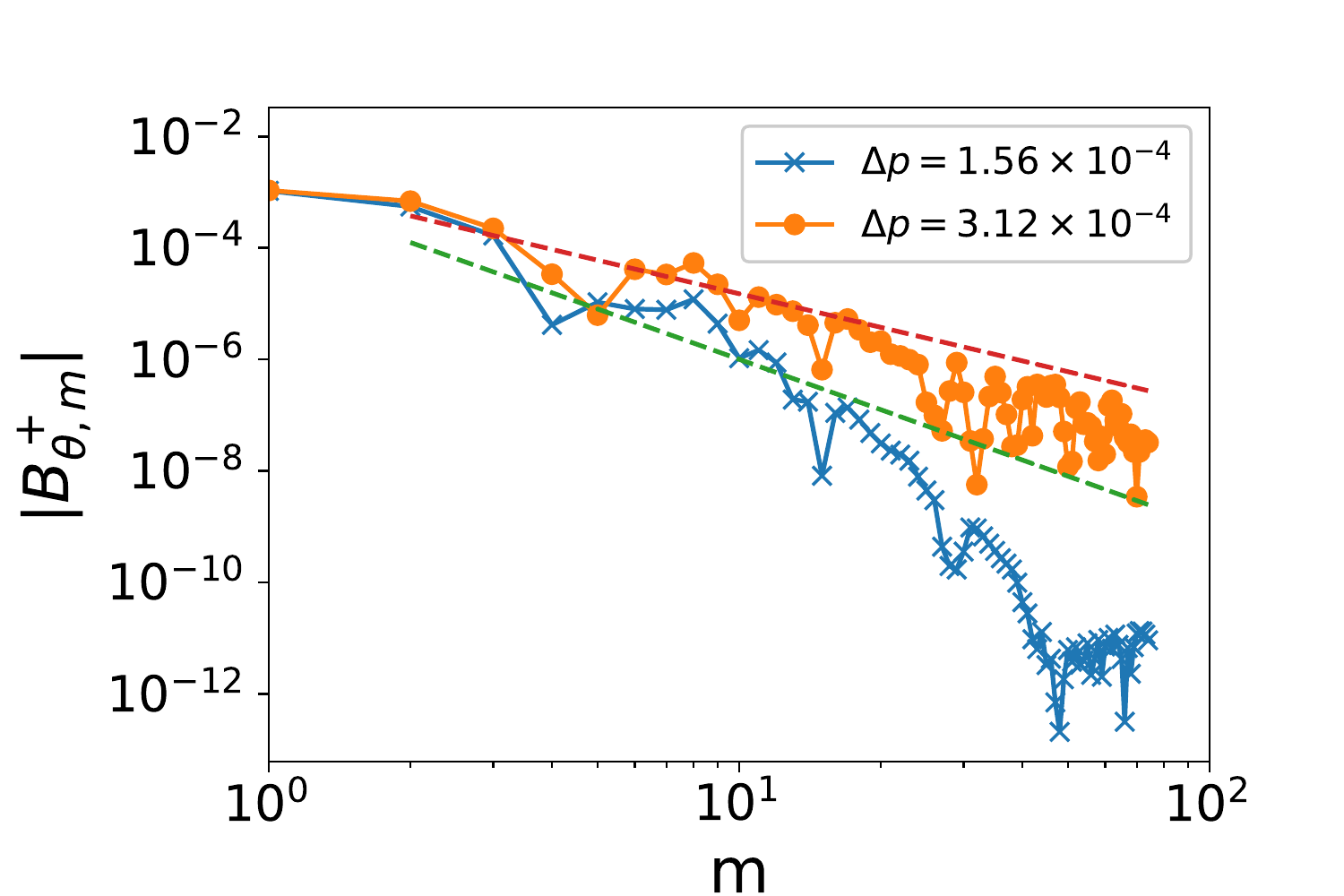}
    \caption{\modi
    The Fourier harmonics $|B_{\theta,m}^+|$ of $B_\theta^+(\theta, 0)$,
  on the interface between the second and third volume the three-volume HKT equilibrium,
  for $\delta = 0.014$ and two different $\Delta p$ (solid lines), 
  computed with $M=77$, $N=30$, and $L=10$. 
  The harmonics are plotted in a log-log scale with $m$ being the poloidal Fourier mode number.
  The upper and lower dashed lines show trend lines of $m^{-2}$ and $m^{-3}$, respectively.
  \norm}
    \label{fig:Btheta_delta014}
\end{figure}

The transition from the existence of an analytic solution to non-existence can be clearly seen from the Fourier spectrum of the interface.
This is demonstrated by \figref{fig:Rm_delta014},
in which an increase in the pressure jump results in a loss in the exponential decay in the tail of the Fourier series.
\modi
According to \Figref{fig:alpha_pressure},
the case with $\Delta p = 3.12 \times 10^{-4}$ sits on the boundary of existence/non-existence.
Its Fourier harmonics decay as $m^{-3}$ instead of $m^{-2}$ in the case of a boundary circle.
Even though the interface is non-analytic, one cannot claim directly the non-existence of solution from the smoothness class of the interface like in the zero-pressure case.
Instead, we have plotted the Fourier harmonics $B_{\theta,m}^+$ of $B_\theta^+(\theta,0)$ in \figref{fig:Btheta_delta014}.
The figure clearly show that $B_{\theta,m}^+$ is bounded between the trend lines of $m^{-2}$ and $m^{-3}$, just like that of a boundary circle in the real space.
This indicates that the KAM surface containing the solution to the force balance problem is now critical.
A further small increase in $\Delta p$ will lead to the non-existence of solution.
Therefore, the non-existence of an analytic solution here also means the non-existence of a solution in general.
\norm

\modi
In other words, a higher pressure jump pushes the interface closer to being non-analytic,
and thus decreases the maximum pressure jump the interface can support,
according to the result in \secref{sec:PJH_McGann}.
The critical pressure jump is reached when the predicted maximum $\Delta p$ from the PJH matches the exact $\Delta p$ on the interface.
That is, when the KAM surface in the PJH phase space becomes a boundary circle.
A comparison with \figref{fig:allowed_pressure_jump_McGann} identifies the change of interface geometry as a more important effect than the increase of $\Delta p$ in the PJH itself,
as it significantly reduces the allowed pressure jump when taken into account.
\norm

\begin{figure}[htbp]
  \centering
  \includegraphics[width=8cm]{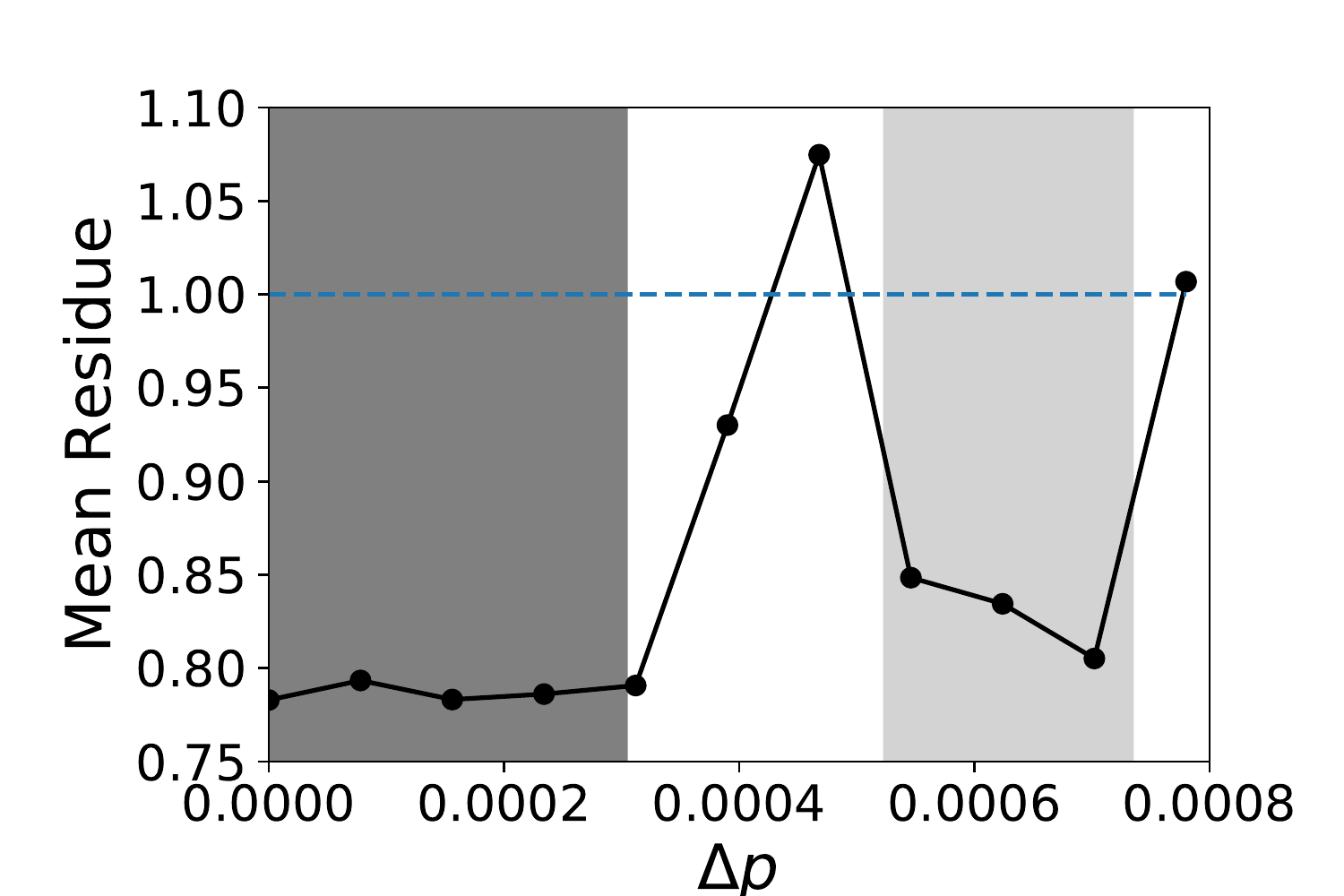}
  \caption{Greene's mean residue \modi of \norm the tenth convergent as a function of $\Delta p$, for $\delta=0.014$.
  The horizontal dash line indicates the threshold greater than which the KAM surface is broken.
  The grey and light grey region are the predicted first and second existence parameter region as shown by \figref{fig:allowed_pressure}.
  The SPEC equilibria are computed with a resolution of $M=20, N=8$ and $L=10$.
  }
  \label{fig:mean_residue}
\end{figure}

It is worthwhile to cross-validate our result with Greene's residue in the PJH phase space.
We will take $\delta = 0.014$ as an example since it contains both existence and non-existence regions.
\Figref{fig:mean_residue} shows Greene's mean residue $\beta_{10}$ (the tenth convergent) of the potential KAM surface in the PJH phase space.
An introduction of the mean residue is given at the end of \ref{app:greenes_residue}.
A mean residue greater than unity indicates the non-existence of solution.
We find that the two non-existence parameter regimes shown by \figref{fig:allowed_pressure} are consistent with the two peaks of the mean residue,
although the intervals of $\Delta p$ given by $\beta_{10}>1$ are narrower than that of \figref{fig:allowed_pressure}.
This is because the residue is underestimated with a finite Fourier resolution, as discussed in \secref{sec:PJH_McGann}. 

\section{Conclusion}
\label{sec:conclusion}
In summary, 
\modi
a solution to the stepped-pressure equilibrium equations does not always exist.
\norm
If the departure from symmetry is large, an interface tends to break into fractal cantori, leading to 
\modii
a fractal interface and thus the non-existence of solution.
\norm
A SPEC numerical solution with a given resolution can still be found,
but the numerical error does not reduce at the expected exponential rate as the resolution increases. 
\modi
To identify converged solutions with analytic interfaces, 
\norm
one can run SPEC at an extremely high Fourier resolution and make sure the Fourier spectrum of the interface has an exponentially converging tail.
Alternatively, one can run the same equilibrium multiple times with different Fourier resolutions and study their convergence.
In this paper, we proposed a criterion based on the convergence property of the return map of the phase space of the pressure jump Hamiltonian (PJH).

The solution to the force balance condition corresponds to a KAM surface in the phase space of the PJH.
This phase space becomes more chaotic as the pressure jump (equivalent to the energy of the Hamiltonian) increases when the interface geometry is kept fixed.
At the maximum pressure jump, the KAM surface is broken, leading to the non-existence of solution.
An interface closer to critical can support a smaller pressure jump, consistent with the findings by McGann. 
Using the convergence criterion we developed,
we discovered that the allowed pressure jump on an interface decreases gradually to zero as the interface approaches a non-smooth boundary circle.
At the same time, a pressure jump on the interface also significantly modifies the force balance,
pushing the interface closer to \modi being non-analytic, \norm and thus reduces the maximum pressure jump it can support.
\modi
Furthermore, we found that sometimes with a higher pressure jump the solution will come back to existence, namely the pressure healing effect.
\norm

A generalisation of our non-existence study to toroidal geometry is straightforward.
We believe the mechanism and behaviour of the transition from solution existence to non-existence is common between the slab and the toroidal geometry.
A notable difference is that our boundary perturbation to the slab only includes three Fourier harmonics,
while in tokamaks with boundary perturbations and stellarators there are more Fourier mode numbers in both poloidal and toroidal directions.
This requires much higher resolution SPEC calculations to show the exponentially converging tail of the Fourier harmonics,
or to perform a convergence scan.

In future works, we propose to develop an interface pruning algorithm that removes an interface when it becomes non-smooth,
or when it tries to support a pressure jump that leads to the solution non-existence.
After the interface is removed, the pressure will be redistributed among neighbouring volumes until a solution is reached.
\modi
By doing so, we can infer the maximum pressure a machine can support,
i.e. the machine $\beta$ limit.
\norm
Another target application is a sawtooth cycle~\cite{Wesson1986} or an edge localised mode (ELM)~\cite{Snyder2014}, 
in which the symmetry-breaking perturbation is provided by the instability and a global redistribution of pressure is reached at the final equilibrium.


\ack
The authors are grateful to the SPEC user/developer group (SPECtaculars) and the collaborators of Simons Collaboration on Hidden Symmetries and Fusion Energy (HSFE) for useful discussions.
In particular we thank D. Pfefferl\'e for a discussion regarding the boundary condition with an X point.
We would like to thank Prentice Bisbal and Caoxiang Zhu for their help to run the high resolution SPEC cases.
This research was undertaken with the assistance of resources and services from the National Computational Infrastructure (NCI), which is supported by the Australian Government.
This work was supported by a grant from the Simons Foundation/SFARI (560651, AB). 
This work is partly funded by Australian Research Council project DP170102606 (ZSQ, RLD, MHJ) and by DOE Contract No. DEAC02–76CH03073 (SRH).
This work has also been
carried out within the framework of the EUROfusion Consortium
and has received funding from the Euratom research and training
programme 2014–2018 and 2019–2020 under Grant Agreement
No. 633053 (JL). The views and opinions expressed herein do not
necessarily reflect those of the European Commission.

\appendix

\section{Greene's residue criterion and its application to the one-volume HKT problem}
\label{app:greenes_residue}
The threshold of $\delta$ that breaks the last standing KAM surface can be estimated by Greene's residue criterion.
In his 1979 paper \cite{Greene1979}, Greene put forward a conjecture based on his numerical experiments with the standard map.
The conjecture was later proved by MacKay \cite{MacKay1992} in some cases.

The conjecture says that the break up of an invariant torus with an irrational rotational transform $\iotabar$ is associated with 
the transition of the periodic orbits with rotational transform $p_i/q_i$ from elliptic to hyperbolic as $i \rightarrow \infty$.
The sequence of coprime integers $p_i/q_i$ are constructed such that $|\iotabar - p_i/q_i| \rightarrow 0$ as $i \rightarrow \infty$. 
The elliptic periodic orbits are the O points of the islands and the transition of them from elliptic to hyperbolic means the loss of stability and the disappearance of the O points.
Greene's conjecture identifies the steps to locate the critical $\delta$ leading to the break up of the KAM surface with transform $\iotabar$,
by examining the stability of its nearby island O points. The steps are as follows.
\begin{enumerate}
    \item Find a convergent sequence $p_i/q_i$ such that $p_i/q_i$ is a better and better approximation of the irrational number $\iotabar$.
    \item Locate the island O points with the rotational transform $p_i/q_i$ as $i \rightarrow \infty$.
    \item Determine $\delta$ for which the island O point disappears.
\end{enumerate}
We will go through this process in detail below.

The convergent sequence $p_i,q_i$ can be constructed by truncating the \textit{continued fraction expansion} of $\iotabar$,
introduced in \eqref{eq:continued_fraction}.
By truncating the continued fraction expansion up to the $i$-th term, we can find the sequence $p_i/q_i = \lcontfrac a_0, \cdots, a_i \rcontfrac$ that satisfies the required property of Greene's conjecture.
The next step is to locate the island O points with rotational transform $p_i/q_i$ and find out their stability.
Let $(s,\theta,\zeta)$ be the coordinate parametrisation of the slab.
The radial coordinate $s$ is introduced such that $s=-1$ and $s=1$ on the lower and upper plasma boundary, respectively.
Given $B^\zeta \ne 0$, the field line equations are given by
\begin{equation}
    \frac{d s}{d \zeta} = \dot{s}(s,\theta,\zeta) = \frac{B^s}{B^\zeta}, \quad
    \frac{d \theta}{d \zeta} = \dot{\theta}(s,\theta,\zeta) = \frac{B^\theta}{B^\zeta}.
    \label{eq:field_line_flow}
\end{equation}
The field line equations are a system of ordinary differential equations.
Given an initial condition, the trajectory of a field line (also named as an \textit{orbit}) can be solved by a standard numerical integrator.
Starting from a given point $s, \theta$ on the $\zeta=\zeta_0$ plane,
one can integrate \eqref{eq:field_line_flow} until $\zeta=\zeta_0+2\pi$ so the field line reaches back to the original plane due to the periodicity in $\zeta$.
This gives the \textit{return map} or \textit{\Poincare map} $F:\real^2 \rightarrow \real^2$ written as,
\begin{align}
    \left(
    \begin{array}{c}
      s' \\
      \theta’
    \end{array}
  \right) = 
  F 
  \left(
    \begin{array}{c}
      s \\
      \theta
    \end{array}
  \right).
  \label{eq:return_map}
\end{align}
We note that $F$ is a twist map since $\iotabar$ is monotonically increasing.
The \Poincare plots in \figref{fig:poincare_1vol}  were constructed by iterating the return map. 

A periodic orbit with rotational transform $p/q$ is a field line that closes back onto itself after $q$ iterations of the return map and rotates a $\theta$ angle of $2 \pi p$.
That is to say
\begin{align}
    \left(
    \begin{array}{c}
      s \\
      \theta + 2 \pi p
    \end{array}
  \right) = 
  F^{q} 
  \left(
    \begin{array}{c}
      s \\
      \theta
    \end{array}
  \right).
\end{align}

The behaviour of orbits in the neighbourhood of an given orbit is described by the tangent map
\begin{align}
    \left(
    \begin{array}{c}
      \delta s' \\
      \delta \theta'
    \end{array}
  \right) = 
  \matT
  \left(
    \begin{array}{c}
        \delta s_0 \\
        \delta \theta_0
    \end{array}
  \right),
\end{align}
in which $\matT$ is a $2\times 2$ matrix
\begin{align}
    \matT = \left( 
        \begin{array}{c c}
            \partial_s s' & \partial_\theta s' \\
            \partial_s \theta' & \partial_\theta \theta' 
          \end{array},
    \right)
\end{align}
constructed by differentiating the return map $F$.
In practice, one needs to solve \eqref{eq:return_map} in addition to \eqref{eq:field_line_flow}. 

The stability of a periodic orbit is related to its tangent map $\matT^{q}$,
which is constructed by iterating the tangent map $q$ times until the orbit closes back to itself.
If the eigenvalues of $\matT^{q}$ are complex conjugates, 
the tangent orbits will display elliptical motion under the mapping near the periodic orbit.
Such a periodic orbit is called \textit{stable} (\textit{elliptic}), corresponding to an O point of an island.
If the eigenvalues are real reciprocals, the tangent motion will either exponentially grow or decay and the periodic orbit is \textit{unstable} (\textit{hyperbolic}).
An elliptic orbit can become hyperbolic if the boundary perturbation is large enough.
This corresponds to the disappearance of an island O point into chaos.

Greene \cite{Greene1979} introduced a quantity $R_G$ (Greene's residue)
which characterizes the stability of periodic orbits converging to the KAM surface and is given by
\begin{equation}
    R_{G,i}=  \frac{2 - \trtrace (\matT^{q_i})}{4},
\end{equation}
in which $\trtrace$ is the trace of the matrix.
For a given KAM surface, the residue is a function of the perturbation $\delta$. 
In his numerical experiments, Greene found that if $R_{G,i}$ approaches zero as $i$ approaches infinity, 
then that irrational surface will exist. 
If, however, $R_{G,i}$ diverges to infinity, then that irrational surface has been destroyed. 
The threshold value is $\lim_{i\rightarrow \infty} R_{G,i} = 0.25$, 
which indicates the KAM surface is critical and any infinitesimal increase in $\delta$ will break it.
In practice, it is increasingly difficult to locate periodic orbits as $q_i$ increases.
To give an estimation of the breakup threshold, the usual process is to terminate the calculate at a certain level, e.g. $i=10$.
For a KAM surface close to breakup, $R_{G,i}$ oscillates around $0.25$ and converges slowly as $i$ increases, making the estimation less accurate.

Greene's residue $R_{G,i}$ quickly converges to zero or diverges to infinity as $i$ increases if the KAM surface is not in the proximity of breaking up.
Greene also defined the \textit{mean residue}, given by
\begin{equation}
  \beta_i = \left( \frac{|R_{G,i}|}{0.25} \right)^{1/q_i},
\end{equation}
which is a less stiff function of the scanning parameter, e.g. the size of the perturbation.
The threshold is now $\beta_i = 1$ for critical,
$\beta_i < 1$ for sub-critical and $\beta_i > 1$ for over-critical.
It is noteworthy that such a criterion can be generalised to the hyperbolic fixed points \cite{Greene1979},
which have $R_{G,i} < 0$.

\begin{figure}[htbp]
  \centering
  \includegraphics[width=8cm]{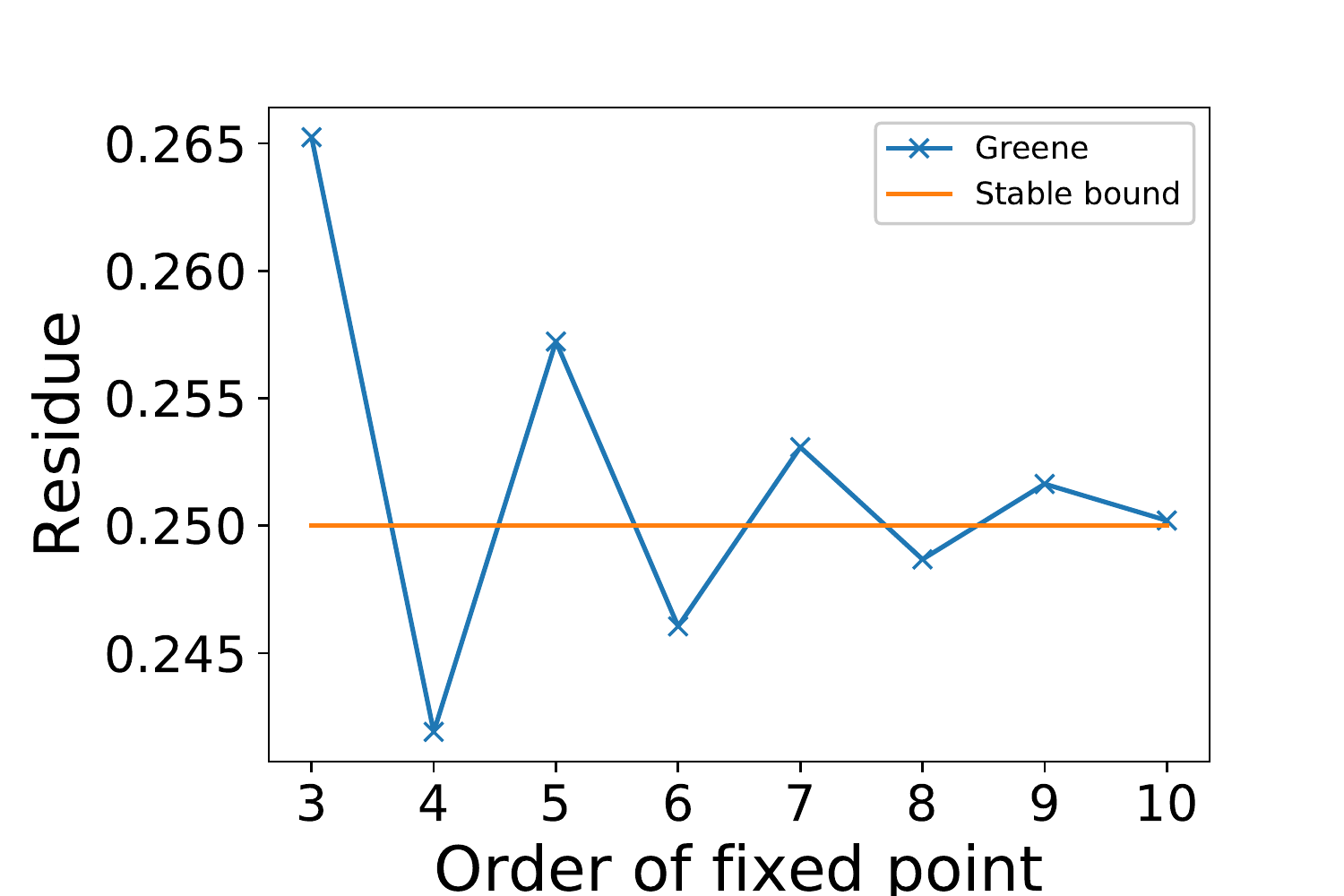}
  \caption{The Greene's residue $R_{G,i}$
  for $\delta = \delta_c =0.020782$ and the KAM surface with $\iotabar = \varphi^{-2} \approx 0.382$. }
  \label{fig:greenes_residue}
\end{figure}

Coming back to the HKT problem in \secref{sec:one_volume}, we are trying to find the break-up threshold for the KAM surface with $\iotabar = \varphi^{-2} \approx 0.382$.
To do that, we write the continued fraction expansion of $\varphi^{-2}$ to be $\varphi^{-2} = \langle 0, 2, 1, 1, cdots \rangle$.
Truncating the first $n$ terms of the continued fraction expansion, we get a series of rational numbers:
$1, 1/2, 1/3, 2/5, 3/8, 5/13, 8/21, 13/34, 21/55, 34/89, \cdots$.
The tenth convergent is $34/89$ and the location of the fixed points with $p/q=34/89$ can be computed using a Newton's method.
The break up threshold $\delta_c$ is found
using a bisection method on the Greene's residue $R_{T,10}$ by iterating $\delta$ until $R_{T,10}=0.25$.
The break-up threshold is estimated to be $\delta_c =0.020782$ with the residue plotted in \figref{fig:greenes_residue}.

\section{Smoothness property of a critical KAM surface}
\begin{figure}[htbp]
  \centering
  \[
    \begin{array}{c c}
      \includegraphics[width=8cm]{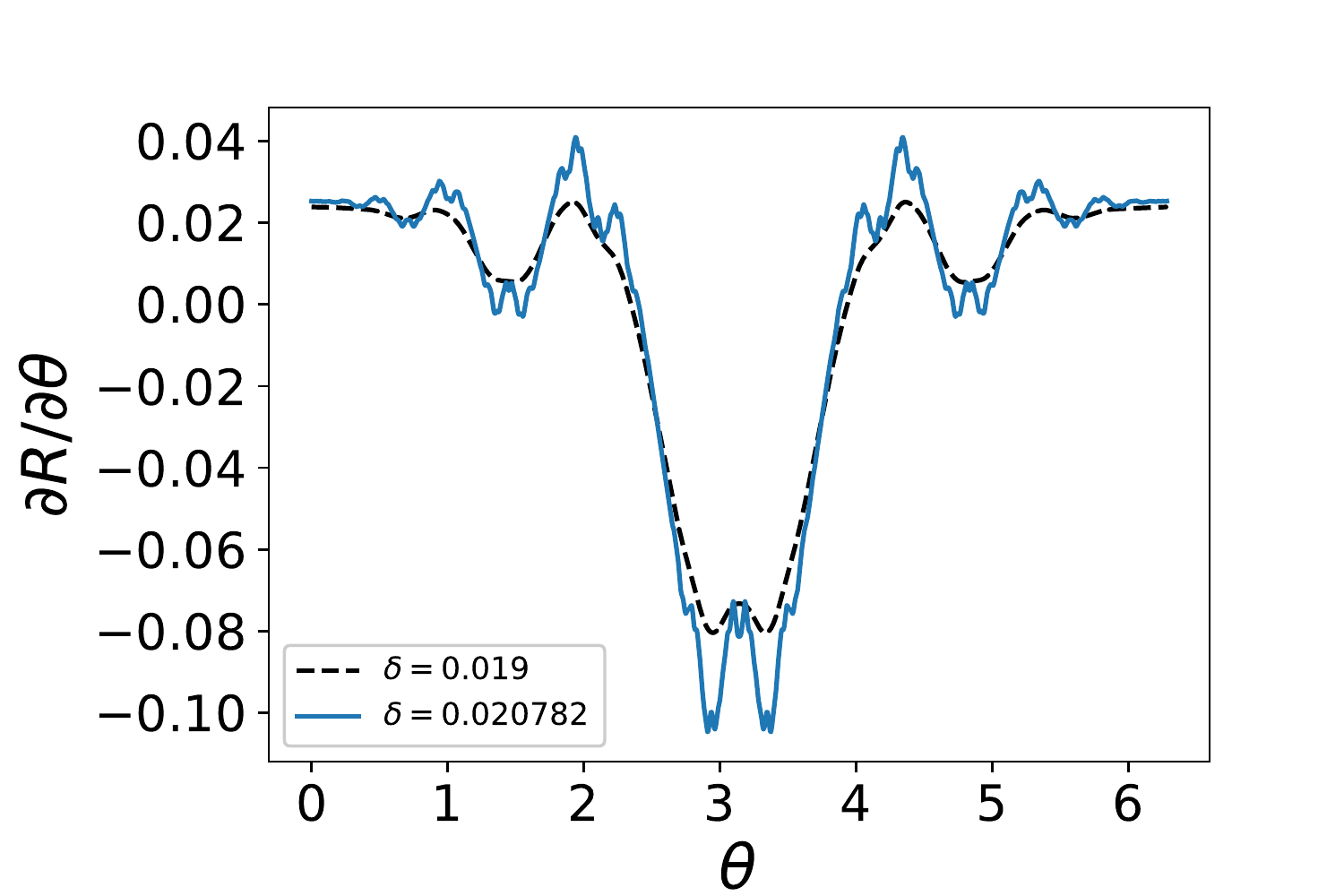} & 
      \includegraphics[width=8cm]{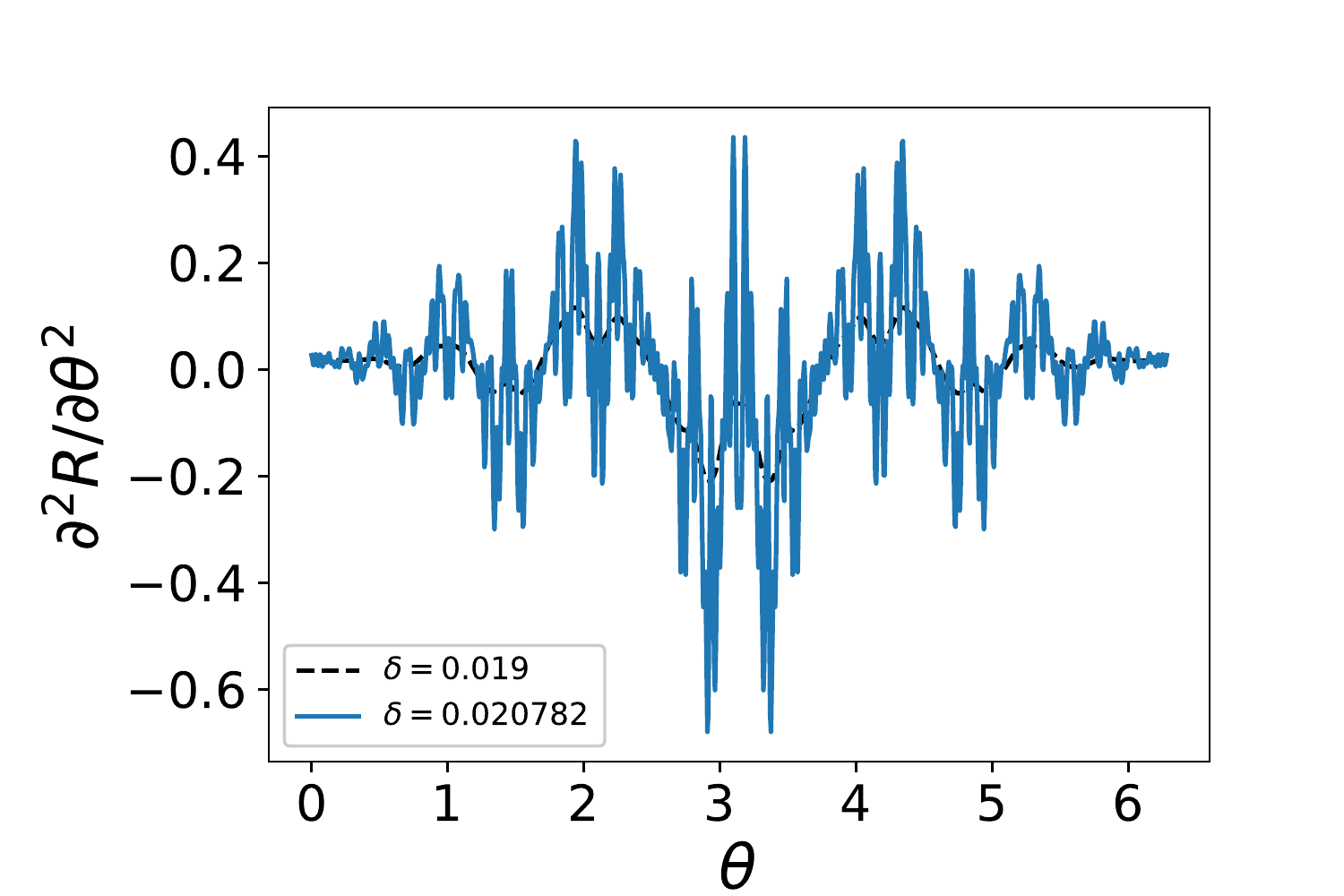} \\
      \text{(a)} & \text{(b)}
    \end{array}
  \]
  \caption{The first (a) and the second (b) $\theta$ derivatives of the KAM surface with $\iotabar=\varphi^{-2}$ in the single volume case,
  for it being sub-critical with $\delta=0.019$ and critical with $\delta=0.020782$.}
  \label{fig:derivative_KAM}
\end{figure}

The smoothness property of the KAM surface with $\iotabar=\varphi^{-2}$ in the single volume case can be found by plotting its derivatives.
In \figref{fig:derivative_KAM} we have plotted its first and second $\theta$ derivatives for both $\delta=0.019$ (sub-critical) and $\delta=\delta_c$ (critical).
The KAM surface is represented by fitting it with 200 Fourier harmonics.
For the critical case, the second $\theta$ derivative has spikes almost everywhere.
Each spike corresponds to a discontinuity in the first derivative, which will generate a $\delta$ function when differentiated once.
Therefore, we conclude that for the KAM surface being a boundary circle,
its first derivative is discontinuous and the function class is $C^0$.
To the contrary, both the first and second derivatives of the sub-critical case are smooth.

\label{app:KAM_smoothness}

\bibliography{references}

\end{document}